\documentclass[useAMS,usenatbib,usegraphicx]{mn2e}

\def\lya{Ly-$\alpha$}

\def\nhi{\rm{N}$_{\rm{H\,{I}}}$}
\def\kms{km s$^{-1}$}

\title[The association between gas and galaxies]{The association
between gas and galaxies I: CFHT spectroscopy and pair analysis}
\author[Simon L. Morris and Buell T. Jannuzi]{Simon
L. Morris$^{1}$\thanks{Visiting Astronomer, Canada-France-Hawaii
Telescope operated by the National Research Council of Canada, the
Centre National de la Recherche Scientifique de France and the
University of Hawaii.} and
Buell T. Jannuzi$^{2}$\footnotemark[1]\\
$^{1}$E-mail: Simon.Morris@Durham.ac.uk, Department of Physics,
University of Durham, South Road, Durham, DH1 3AJ, UK.\\
$^{2}$E-mail: jannuzi@noao.edu, National Optical Astronomy
Observatory, PO Box 26732, Tucson, AZ 85726-6732, USA.}
\begin{document}

\date{Accepted . Received ; in original form }

\pagerange{\pageref{firstpage}--\pageref{lastpage}} \pubyear{2005}

\maketitle

\label{firstpage}

\begin{abstract}

We investigate the relative distribution of the gaseous contents of
the Universe (as traced by a sample of Lyman $\alpha$ (\lya\ )
absorbers), and the luminous baryonic matter (as traced by a
redshift survey of galaxies in the same volume searched for \lya\
absorbers), along 16 lines-of-sight (LOS) between redshifts 0 and 1.
Our galaxy redshift survey was made with the Multi-Object
Spectrograph (MOS) on Canada-France-Hawaii Telescope (CFHT) and,
when combined with galaxies from the literature in the same LOS,
gives us a galaxy sample of 636 objects. By combining this with an
absorption line sample of 406 absorbing systems drawn from published
works, we are able to study the relationship between gas and
galaxies over the latter half of the age of the Universe. A
correlation between absorbers and galaxies is detected out to
separation of 1.5 Mpc. This correlation is weaker than the
galaxy-galaxy correlation. There is also some evidence that the
absorbing systems seen in CIV are more closely related to galaxies,
although this correlation could be with column density rather than
metallicity. The above results are all consistent with the absorbing
gas and the galaxies co-existing in dark matter filaments and knots
as predicted by current models where the column density of the
absorbing gas is correlated with the underlying matter density.

\end{abstract}


\section{Introduction} \label{sec-intro}

This paper is part of our efforts to provide constraints and
measurements of the relationship between the distribution of the
gaseous contents of the Universe as traced by neutral Hydrogen and
the luminous baryonic mater as traced by galaxies.

The manner in which gas collapses gravitationally into dark matter
potential wells to form stars (and hence galaxies), and the way in
which these stars then affect the gas is a topic of great interest
at present. This is being pursued both observationally and
theoretically at redshifts from 7 to zero (i.e. when the universe
was approximately 5\% of its current age, to the present day). In
this paper we investigate this process observationally during the
second half of the evolution of the universe.

There is a continuing debate about the relationship between low
redshift Lyman~$\alpha$ (\lya) absorbers and galaxies, where here
`low redshift' is taken to mean redshifts less than 1. A simplified
(strawman) version of the two sides is (a) that all low redshift
\lya\ absorbers are part of physically-distinct luminous-galaxy
halos, or (b) that they are all part of the filamentary structure
seen in recent SPH/Mesh structure formation models, and are only
related to galaxies by the fact that both are following the
underlying dark matter distribution.  In practice, almost all
authors acknowledge that the universe includes a mix of the above
two populations (and indeed others), and the debate is more about
which population dominates a particular set of observations.

Both of these positions have been vigorously defended in the
literature.  The reference list below includes all relevant papers
listed on the ADS abstract server from the time of the first
available high quality UV spectroscopy from HST to 15 August 2005.

Papers supporting (a) include (in chronological order):
\citet{1994MNRAS.269L..49M}; \citet{1995ApJ...442..538L};
\citet{1996ApJ...456L..17L}; \citet{1998Ap&SS.263...75B};
\citet{1998ApJ...495..637L}; \citet{1998ApJ...498...77C};
\citet{1999ApJ...523...72O}; \citet{2000ApJ...529..644L};
\citet{2001ApJ...556..158C}; \citet{2001ApJ...559..654C};
\citet{2002ApJ...570..526S}; \citet{2003ApJ...589..111C};
\citet{2004ApJ...606..196Z}; \citet{2004ApJ...609...94S};
\citet{2004MNRAS.354L..25B}; \citet{2005A&A...429L...5D};
\citet{2005ApJ...620...95T}; \citet{2005ApJ...622..267K};
\citet{2005ApJ...623...57M} and \citet{2005ApJ...623..767J}.

Papers supporting (b) include (in chronological order):
\citet{1991ApJ...377L..21M}; \citet{1993ApJ...419..524M};
\citet{1994ApJ...427..696M}; \citet{1994MNRAS.269...52M};
\citet{1995ApJ...438..650W}; \citet{1995Natur.373..223D};
\citet{1995ApJ...451...24S}; \citet{1996AJ....111...72S};
\citet{1996ApJ...457...19B}; \citet{1996ApJ...458..518R};
\citet{1996A&A...306..691L}; \citet{1996ApJ...464..141B};
\citet{1996AJ....112.1397V}; \citet{1997ApJ...491...45D};
\citet{1998ApJS..118....1J}; \citet{1998ApJ...505..506G};
\citet{1998ApJ...506....1W}; \citet{1998ApJ...508..200T};
\citet{1999ApJS..122..355V}; \citet{1999ApJ...524..536I};
\citet{2000ApJ...544..150P}; \citet{2000ApJS..130..121P};
\citet{2002ApJ...565..720P}; \citet{2002ApJ...575..697T};
\citet{2002ApJ...575..697T}; \citet{2002ApJ...574L.115M};
\citet{2002ApJ...574..599M}; \citet{2002ApJ...574L.115M};
\citet{2002ApJ...580..169B}; \citet{2003ApJ...591..677R};
\citet{2003ApJ...591...79M}; \citet{2004ApJS..152...29P};
\citet{2004ApJ...614...31B}; \citet{2004ApJS..155..351S};
\citet{2005ApJ...618..178C}; \citet{2005ApJ...624..555D} and
\citet{2005ApJ...629L..25C}.

Finally to complete the reference list, some of the theoretical
papers from the last ten years which are relevant for this debate
are: \citet{1996ApJ...457L..51H}; \citet{1997ApJ...485..496Z};
\citet{1998ApJ...496..577C}; \citet{1998MNRAS.297L..49T};
\citet{1999ApJ...511..521D}; \citet{1999ApJ...514....1C};
\citet{2001ApJ...553..528D}; \citet{2001ApJ...559..507S};
\citet{2002ApJ...580...42M}; \citet{2002ApJ...574..590S};
\citet{2002MNRAS.336..685V}; \citet{2002ApJ...578L...5T};
\citet{2002ApJ...580...42M}; \citet{2004MNRAS.348..421N};
\citet{2004MNRAS.348..435N}; \citet{2004ApJ...613..159F};
\citet{2005ApJ...620L..13A}; \citet{2005ApJ...623L..97T};
\citet{2005ApJ...624L...1S}; \citet{2005MNRAS.359..295F};
\citet{2005MNRAS.360.1110V} and \citet{2005MNRAS.361...70J}.

Regardless of whether either hypothesis is correct, we emphasize
that even if one (naively) believes a clear answer can be obtained
(either (a) or (b) above), this answer will be a function of the
neutral Hydrogen column density of the absorbers.  For
\nhi$>10^{21}$ cm$^{-2}$, one is probing a column density of
material comparable to that of the disk of our galaxy, and hence all
galaxies, to varying degrees, should be expected to contribute to
the population of such absorbers. At column densities
\nhi$\sim10^{12}$ cm$^{-2}$, one is getting close to the neutral
hydrogen content of the expected fluctuations within voids, and we
expect that a more heterogeneous set of causes could produce such
absorption.

The increasingly sophisticated models of galaxy formation and
evolution now available (see references above) suggest that if we
could make a perfect census of the gaseous and luminous constituents
of a large volume of the low redshift Universe, we would find that
both the gas (quasar absorption line systems) and stars (galaxies)
trace the same fundamental structures (whose course of formation and
distribution was set by the early perturbations of the distribution
of dark matter in the Universe). These same models, however, also
indicate that a wide range of detailed relationships between the gas
and galaxies should be observed.  The details of the structures will
depend on an equally wide range of interesting astrophysics,
including the process of star formation and ``galaxy feedback'' and
we hope that the approach of this paper can lead to a better
understanding of this astrophysics.

\begin{figure}
\includegraphics[height=70mm, angle=0]{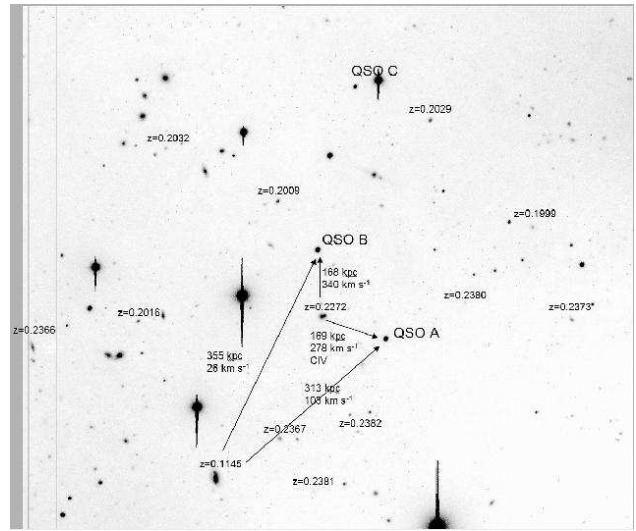}
 \caption{Illustrative image of the two lines-of-sight (LOS) to the
 QSOs LBQS 0107-025A and B. The region shown is 7.1'$\times$6.2', with
 North up and East to the left. The two QSOs discussed in this paper
 are labelled, as is a third QSO (labelled C) to the North
 \citep{2001ApJ...549...76Y}. A
 few of the measured galaxy redshifts
 in the field are labelled (drawn so the decimal point is approximately
 above the relevant galaxy) and the projected spatial offset and
 velocity difference to detected absorbers in either of the LOS are
 marked for two example galaxies. See the text for a discussion.
 The image shown was made from
 a 90 second exposure without a filter using the Palomar Observatory
 5m telescope and COSMIC imager/multi-object spectrograph on September
 22, 1995UT.}
  \label{fig-q0107_lab}
\end{figure}

Observationally comparing the distribution of galaxies and quasar
absorption line systems shows some of the complexity of the
phenomena at work. This can be illustrated by considering
Figure~\ref{fig-q0107_lab}.  In this figure we show an image of the
lines-of-sight (LOS) towards the QSOs LBQS 0107-025A and B. As
discussed later in the paper, for this field we have samples of
galaxies and \lya\ absorption line systems whose locations in this
common volume of space can be compared.  In the figure we have
selectively labelled illustrative cases of `associated' galaxies and
absorption lines. For two of the galaxies labelled with a redshift,
arrows indicate where there is a detected absorption line in both of
the QSO spectra. For those absorber-galaxy pairs, the projected
spatial separation and velocity difference between the galaxy and
the absorber are also provided.

We consider a few of these cases below:
\begin{enumerate}
\item A bright, multi-component object lies between the two QSOs at
a redshift of 0.2272. There is absorbing gas seen at nearly this
same redshift in both QSO LOS, and in one of them, CIV absorption is
also detected. This then could potentially be interpreted as a
straightforward case of a large gaseous halo around a bright galaxy.
The complex morphology of the galaxy could indicate an ongoing
merger.
\item At a distance of 350 kpc to the south and east of the two QSO LOS is another
bright morphologically-complex galaxy at a redshift of 0.1145. This
galaxy is also close in velocity to absorption seen in both LOS. It
is instructive though to compare the small visible extent of the
stars in the (large) galaxy with the distance to the two QSO LOS.
Smooth, undisturbed gaseous halos of this size seem improbable,
although it is one of the goals of this paper to investigate this
statistically.
\item Generally to the north of the two QSOs are a collection of 5 galaxies
marked with redshifts near 0.20. All 5 of these galaxies are within
500 \kms\  (and also within a projected distance of 500 kpc) of an
absorber seen in the LOS to LBQS 0107-025B. No absorption is
detected in the other QSO LOS. This collection illustrates the
common difficulty of picking out one individual galaxy to tie to one
absorber. Although there is indeed a single `closest' galaxy, it
seems rather arbitrary to claim that it is the sole cause of any
absorption.
\item Finally, a collection of six galaxies, all within 500 km
s$^{-1}$ of each other at a redshift around 0.24 are marked
surrounding the LOS to LBQS 0107-025A. Despite their proximity to
the LOS, and the case above where a group of galaxies can be
associated with an absorber, no absorption is seen at this redshift
in either of the two QSO LOS. While one could speculate that low
column density absorbers might be fragile entities that are
destroyed in a dense environment, there is no good independent
evidence at present that this galaxy `group' has crossed such a
threshold.
\end{enumerate}

These examples demonstrate the great variety of appearances the
underlying relationship between the absorbers and galaxies presents,
along even one LOS. This richness should be kept in mind when
interpreting the results derived from the ensemble of data we
consider in this paper.

The outline of the paper is as follows: in \S~\ref{sec-data} we
describe the various data sets used in this paper. In
\S~\ref{sec-anal} we analyse the data, and in \S~\ref{sec-conc} we
present our conclusions.

\section{Data}
\label{sec-data}

Our observational program would ideally be designed to allow a
statistical comparison between the absorption gas detected in Hubble
Space Telescope (HST) UV spectroscopy between redshifts of 0 and 1
and the galaxies within a cylindrical region at least 3
Mpc\footnote{Throughout we assume H$_0$=70 km s$^{-1}$ Mpc$^{-1}$,
$\Omega$=0.3 and $\Lambda$=0.7. Impact parameters are reported in
physical rather than comoving coordinates.} in radial extent away
from the QSO Line of Sight (LOS). In practice, what we achieved was
an absorption line sample taken from the HST Quasar Absorption Line
Survey \citep{1993ApJS...87....1B, 1996ApJ...457...19B,
1998ApJS..118....1J}, an HST Key Project during cycles 1-4, with
varying redshift coverage, and galaxy samples from a variety of
observatories. The main one reported here was obtained using the
Multi-Object Spectrograph (MOS) at the Canada-France-Hawaii
Telescope (CFHT). Some additional galaxy redshifts have been
obtained using the COSMIC spectrograph at Palomar, and gleaned from
the literature. We are in the process of supplementing this data set
with additional observations using other facilities. Although we do
not claim that any of these galaxy samples are `complete', we have
endeavoured to ensure that they all come with known selection
functions, allowing a statistical analysis to be performed.
Completeness in this context would be rather an illusory concept, as
there will always be weaker lines and less luminous or lower surface
brightness galaxies that are missed by any survey. This point has
been made by many authors including \citet{1998ApJ...495..637L,
2000ApJ...529..644L} and \citet{ 2003ApJ...591..677R}.

\subsection{Sample definition} \label{subsec-sample}

The fields observed with MOS on the CFHT were chosen to include Key
Project QSOs \citep{1998ApJS..118....1J} with QSO redshift greater
than 0.5 and Faint Object Spectrograph (FOS) higher dispersion
spectroscopy covering from redshift 0.34 to 1 (i.e., FOS gratings
G190H and G270H). While UV spectroscopy down to z=0 is obviously
desirable (and has been obtained with the HST along a limited number
of sight lines) the KP LOS including coverage to z=0 (observations
made with the FOS G130H grating) are of QSOs with redshifts too low
for efficient CFHT MOS followup. In Table~\ref{tab-sample} we list
the QSOs selected. The columns include the QSO RA, Dec, redshift and
V band magnitude, followed by the date of the HST UV spectroscopy
for the 3 FOS gratings used in this paper, followed by the number of
CFHT MOS masks observed (see \S~\ref{subsec-spect}) and the date of
the MOS observation.

There is one additional field observed as part of the CFHT MOS
sample which was not part of the HST Key Project. HST UV spectra of
the QSO pair LBQS 0107-025A and LBQS 0107-025B were studied by
\citet{1995Natur.373..223D}, \citet{1997ApJ...491...45D} and
\citet{2001ApJ...549...76Y}. This LOS was included in the CFHT MOS
target list because of the presence of two QSOs bright enough for
reasonable S/N HST UV spectroscopy within the MOS field of view. An
analysis taking full advantage of this will be performed in a later
paper, but for the present paper we generally use them as two
`separate' LOS. We have checked whether removing one of the LBQS
0107-025 QSOs from the sample makes any significant difference to
our numerical results or conclusions, and have found that it does
not.

\begin{table*}
\begin{minipage}{140mm}
\caption{CFHT MOS Sample Definition} \label{tab-sample}
\begin{tabular}{lccccccccccccccc}
\hline
OBJECT         & R.A.(J2000)& Decl.(J2000) & z  & V     &  G190H   & G270H    & G160L    &  Masks & Date \\
\hline
NAB 0024+22... & 00 27 15.4 & +22 41 59 & 1.118 & 16.6  & 07/22/92 & 07/22/92 & 05/11/91 & 1 & Jul-95 \\
LBQS 0107-025A & 01 10 13.1 & -02 19 52 & 0.960 & 17.9  & 02/05/97 & 12/26/96 &          & 1 & Jul-95 \\
LBQS 0107-025B & 01 10 16.2 & -02 18 50 & 0.956 & 17.3  & 02/12/94 & 12/26/96 &          & 1 & Jul-95 \\
PKS 0122-00... & 01 25 28.9 & -00 05 57 & 1.07  & 16.7  & 11/09/91 & 09/11/91 &          & 2 & Nov-97 \\
3C 57...       & 02 01 57.2 & -11 32 34 & 0.67  & 16.4  & 11/01/95 & 11/01/95 &          & 1 & Nov-97 \\
3C 110...      & 04 17 16.7 & -05 53 46 & 0.773 & 15.9  & 03/16/95 & 03/16/95 &          & 3 & Nov-97 \\
NGC 2841 UB3...& 09 19 57.7 & +51 06 10 & 0.553 & 16.5  & 03/04/92 & 03/04/92 & 03/04/92 & 2 & Nov-97 \\
0959+68W1...   & 10 03 06.8 & +68 13 18 & 0.773 & 15.9  & 11/12/92 & 11/12/92 & 11/12/92 & 1 & Nov-97 \\
4C 41.21...    & 10 10 27.5 & +41 32 39 & 0.613 & 16.9  & 09/12/92 & 10/12/92 & 10/12/92 & 2 & Nov-97 \\
3C 334.0...    & 16 20 21.8 & +17 36 24 & 0.555 & 16.4  & 03/04/92 & 03/04/92 & 03/04/92 & 2 & Jul-95 \\
PG 1718+481... & 17 19 38.3 & +48 04 12 & 1.084 & 14.7  & 05/13/93 & 05/13/93 &          & 1 & Jul-95 \\
PKS 2145+06... & 21 48 05.5 & +06 57 39 & 0.99  & 16.5  & 10/22/91 & 10/22/91 &          & 3 & Jul-95 \\
3C 454.3...    & 22 53 57.8 & +16 08 53 & 0.859 & 16.1  & 11/15/91 & 11/15/91 & 11/15/91 & 2 & Jul-95 \\
PG 2302+029... & 23 04 45.0 & +03 11 46 & 1.052 & 15.8  & 05/10/94 & 02/08/94 &          & 2 & Nov-97 \\
PKS 2340-036...& 23 42 56.6 & -03 22 27 & 0.896 & 16.0  & 10/28/91 & 01/18/93 & 01/18/93 & 2 & Nov-97 \\
PKS 2344+09... & 23 46 36.9 & +09 30 46 & 0.677 & 16.0  & 07/10/92 & 07/10/92 & 07/10/92 & 2 & Nov-97 \\
\hline
\end{tabular}
\end{minipage}
\end{table*}

\subsection{HST Quasar Absorption Line Sample} \label{subsec-kpabs}

The complete list of selection criteria for the KP quasars is
discussed in section 2.1 of \citet{1998ApJS..118....1J}.  We review
here only the criteria used by the KP that might produce a bias for
our study. First is their practical decision to select targets with
bright V-band magnitudes and well determined redshifts in order for
their targets to have predictable UV luminosities that, in the
absence of high column density absorption systems along the LOS,
would enable a high signal-to-noise ratio UV spectrum to be obtained
in a modest exposure time with HST. A sample biased toward objects
bright in V-band apparent magnitude might be more inclined to
include lensed quasars, i.e. objects with a foreground over-density
along the LOS. In practice, the redshifts of the quasars included in
the KP are low, and therefore less likely to be lensed. Second, a
subset of the quasars included in the KP sample were not randomly
selected, but were chosen for a GTO program of J.~Bahcall because
they were behind foreground objects whose absorption properties were
to be investigated. These include AGN like Markarian~205 (with the
galaxy NGC~4319 very close to the LOS to Mkn~205, producing
absorption in the spectrum of the background AGN;
\citet{1970ApJ...161L.113W}; \citet{1992ApJ...398..495B}) and 3C~232
(with the galaxy NGC 3067 in the foreground). Of the objects we
selected from the KP catalogues for observation with CFHT, only
NAB~0024$+$22 was not a randomly selected target by the KP. This
quasar was part of the GTO program of Bahcall and was selected
because it was discovered as part of a search for quasars with LOS
near a cluster of galaxies \citep{1973ApJ...183..777B}.  Abell 31,
with a redshift of 0.160, is along the LOS to this quasar.  The
cluster redshift is smaller than the range we investigate along this
LOS.

We had two additional criteria for selecting our sample for the CFHT
observations. The fields needed to be observable during our
scheduled CFHT observing runs. We chose to focus on LOS with quasars
with redshifts greater than 0.5, in order to maximize the efficiency
of our redshift galaxy survey.

With the previously mentioned exception of LBQS~0107$-$025A and B,
the sample of absorption lines we use is drawn from the KP catalogue
papers, \citet{1993ApJS...87....1B}, \citet{1996ApJ...457...19B},
and \citet{1998ApJS..118....1J}. The KP \lya\ absorption line
catalogue was constructed to include all features with a
significance level greater than 4.5$\sigma$ (as defined and
discussed by \citet{1993ApJS...87...45S}). This corresponds to a
variable equivalent width limit along each LOS. Plots showing these
equivalent width limits as a function of wavelength are given in
Figure~2 of \citet{1998ApJS..118....1J}. With the modest spectral
resolution of the FOS (R=1300; approximately 230 \kms), the \lya\
absorption features were generally unresolved and conversion from
the observed equivalent width to column density in neutral hydrogen
must be done with some assumption about the Doppler-b parameter for
the lines. For the subset of lines that were broader than the
instrumental profile, their width probably indicates the blending of
multiple absorption features, and not a line profile that can be fit
to deduce the column density of the absorber. We note that
approximately 10\% of the detected Key Project \lya\ absorbers were
in fact resolved, and, as noted by \citet{1998ApJS..118....1J}, are
a sign of possible clustering of absorbers on these velocity scales.
This results in an additional contribution to the uncertainty in the
derived column densities.

For LBQS~0107$-$025A and B we have chosen to use the line lists
from \citet{2001ApJ...549...76Y}, supplemented with lines from the
GHRS observations of \citet{1997ApJ...491...45D}. (In detail this
meant adopting the lines from Table 5 of
\citet{1997ApJ...491...45D}, and Tables 3 and 5 of
\citet{2001ApJ...549...76Y} for LBQS~0107$-$025A, and from Table 5
of \citet{1997ApJ...491...45D}, and Tables 4 and 6 of
\citet{2001ApJ...549...76Y} for LBQS~0107$-$025B.) We shifted the
wavelengths of the \citet{1997ApJ...491...45D} by -1.72\AA\ as
recommended by \citet{2001ApJ...549...76Y}.

We want the subset of absorbers caused by intervening, rather than
associated, absorption systems. We follow \citet{1998ApJ...506....1W}
and exclude any line from our sample with a difference of less than
3,000 \kms\ from the systemic velocity of the quasar. Note that while
this is a traditional choice of velocity range for excluding
associated systems, it will not exclude all associated systems. In
particular, one of our quasars, PG~2302$+$029, includes at least one
associated system with an ejection velocity of over 50,000 \kms
(\citet{1996ApJ...470L..11J}; \citet{2003ApJ...590...66S}; Jannuzi et
al. 2006, in preparation).  We have excluded the broad system at
$z=0.695$ from the sample used in this paper.

All of the line lists we use have complete identifications with the
exception of the spectrum of PG1718$+$481. The effects of the
incomplete identification of the PG1718$+$481 absorbers are
discussed in \S~\ref{subsec-monte}, where it is shown they are
negligable.

\subsubsection{Comments on Individual Lines of Sight}

Detailed notes about the HST FOS spectra along the individual LOS we
are studying, and the resulting absorption line lists, are included
in the KP catalogue papers and in \citet{2001ApJ...549...76Y}(for
the LBQS~0107$-$025 LOS).  Here we comment on the special properties
of a few of the fields.

As previously mentioned,  NAB~0024$+$22 was discovered during a
search for quasars near or behind clusters of galaxies
\citep{1973ApJ...183..777B}. Abell 31, with a redshift of 0.160,
is foreground to the redshift range covered by the FOS
spectroscopy (0.2950 to 1.0998). Intervening absorption line
systems including absorption by CIV are at $z=0.4069$, 0.4830,
0.8196, and 1.1102.  The SNR of the KP spectrum of this quasar is
not as high as the typically observed object, resulting in a
smaller sample of detected absorbers given the relatively long
redshift path observed ($\delta z=0.8$).

\citet{1991A&A...243..344B} reported an emission line galaxy at
z=0.791 in the LOS to PKS~2145$+$06, which was later identified as
being associated with an extensive metal line systems in the FOS UV
spectrum obtained by the KP \citep{1994ApJ...436...33B}. This
$z=0.791$ system is high enough column density to produce a Lyman
limit system starting at 1633\AA\ , preventing the observation of
\lya\ absorption along this LOS below z$\sim$0.3.

\subsection{Imaging} \label{subsec-image}

To compare the distribution of the galaxies along the same LOS
that the FOS spectroscopy maps the gas distribution requires
knowledge of the positions and redshifts of galaxies in the fields
of our quasar sample.  Images of the quasar fields can provide the
positions and magnitudes of the galaxies. The spectroscopy that
allows the determination of the galaxy redshifts requires the
imaging catalogue as input for the design of multi-slit masks.

For the CFHT MOS, the images used to generate the multi-slit mask
designs are obtained during the same observing block as the
spectroscopy (i.e., the instrument has not been removed from the
telescope between the imaging and spectroscopy observations).
Therefore we obtained images for our fields during our two
observing runs. The images were reduced shortly after they were
obtained and catalogues of galaxies generated to guide the design
of the multi-slit masks.  After the observing runs were completed,
the images were re-reduced, the photometry and astrometry of the
images calibrated, and final catalogues were constructed.  Since
the main purpose of the scheduled observing time was to perform
the spectroscopy, only $R-$band images deep enough to detect the
galaxies we wished to target for spectroscopy were obtained. It
would be desirable to obtain images including additional
band-passes and with enough photometric standard observations to
improve the accuracy of our photometry.

In the following subsections we describe the imaging observations,
the astrometric and photometric calibration of the images, and the
measurement of the galaxy positions and $R-$band magnitudes in the
fields of our targeted quasars.

\subsubsection{Observing}

We had two observing runs with CFHT MOS (July 29 - August 1, 1995 UT
and November 29 - December 3, 1997 UT).  The date each field was
imaged is listed in Table~\ref{tab-imqual}.  The instrument and
telescope properties were significantly different during the two runs
and we will describe the particulars by run.

During the 1995 observing run, the 29th of July UT was photometric
and 600 second $R-$band exposures were obtained of six quasar
fields.  Two and 10 second exposures of the photometric standard
stars in the field of NGC~7790 \citet{1992PASP..104..553O} were also
obtained to allow a photometric zero-point to be determined for the
images. Unfortunately, the proper baffles were left off of the
telescope when MOS was installed on the telescope prior to our run.
As a result, bright arcs of light, produced by stars outside of the
FOV, cut across the images and effectively increased the sky level
over approximately one third of the FOV of the images. The amount
and distribution of the scattered light was, naturally, pointing
dependent.  The mixture of sharp and diffuse features in the
scattered light made it difficult to completely remove during the
processing of the images. This complicated the determination of the
sky background around objects, and the subsequent generation of the
galaxy catalogues necessary for the construction of the multi-slit
masks.  The two most strongly affected images were those of
LBQS~0107$-$025AB and PG~1718$+$481.  To increase the area that
could be mapped for galaxies, we obtained two additional images of
each field, each offset in position from the others, and combined
the images of each field after masking the regions with the worst
scattered light in each of the exposures. The combined images
allowed us to produce improved measurements of the galaxies in these
fields, but there are still regions in each image where we would not
be able to detect galaxies and objects whose measured properties
were strongly affected by scattered light. The imaging on 30 and 31
July 1995 UT was not done under photometric conditions, so the flux
zero-point for the combined image was set to match the imaging on
the 29th.  The CCD in MOS during this run was the Loral-3 with
pixels of approximately 0.313\arcsec\ in size and a gain of 1.9
electrons per ADU.

During the 1997 observing run, the 29th of November UT was
photometric and 600 second exposures of four quasar fields were
obtained. Two and 10 second exposures of the photometric standard
stars in the field of NGC~7790 were also obtained.  The correct
baffles were on the telescope. The nights of UT 26 Nov, 30 Nov,
and 1 Dec, during which 600 second exposures of 5 additional
quasar fields were obtained, were not photometric. For two of the
fields taken during these nights, two 600 exposures were obtained
to compensate. The CCD in MOS during this run was the STIS-2 with
pixels of approximately 0.44\arcsec\ in size and a gain of 4.52
electrons per ADU.

\begin{table}
\caption{Image Quality} \label{tab-imqual}
\begin{tabular}{lccc}
\hline
Object          &  UT Date    &  DIQ     & Comment \\
                &             &  \arcsec &         \\
\hline
NAB 0024+22...  &  29 Jul 95  & 1.75 & \\
LBQS 0107-025AB &  29 Jul 95  & 1.10 & \\
                &  30 Jul 95  & 0.80 &   2 exposures \\
PKS 0122-00...  &  29 Nov 97  & 0.99 & \\
3C 57...        &  30 Nov 97  & 0.85 & \\
3C 110...       &  29 Nov 97  & 1.22 & \\
NGC 2841 UB3... &  26 Nov 97  & 1.23 &   2 exposures \\
0959+68W1...    &  29 Nov 97  & 0.95 & \\
4C 41.21...     &   1 Dec 97  & 1.14 & \\
3C 334.0...     &  29 Jul 95  & 1.15 & \\
PG 1718+481...  &  29 Jul 95  & 0.68 & \\
                &  31 Jul 95  & 0.59 &   2 exposures \\
PKS 2145+06...  &  29 Jul 95  & 0.66 & \\
3C 454.3...     &  29 Jul 95  & 0.88 & \\
PG 2302+029...  &  29 Nov 97  & 0.80 & \\
PKS 2340-036... &  30 Nov 97  & 1.17 & \\
PKS 2344+09...  &  26 Nov 97  & 1.01 &   2 exposures \\
\hline
\end{tabular}
\end{table}

\subsubsection{Reductions}

The images were reduced and analysed twice. During the observing run
a basic reduction was performed with IRAF\footnote{IRAF, the Image
Reduction and Analysis Facility, is a general purpose software
system for the reduction and analysis of astronomical data. It was
written and supported by programmers in the Data Products Program of
the National Optical Astronomy Observatory, which is operated by the
Association of Universities for Research in Astronomy, Inc., under
cooperative agreement with the National Science Foundation.} (bias
subtraction and flat fielding). FOCAS \citep{1989daan.work...35V}
was used to generate a catalogue that could be used for the
identification of galaxies and the generation of the multi-slit
masks. Following the observing runs, the images were processed in a
more relaxed manner, astrometric and photometric calibrations were
determined, and measurements of the object positions and magnitudes
completed. In the following subsections we describe this second
round of reductions, which yielded the measurements presented in
this paper.

The final processing of all of the images was performed using
ccdproc and other tasks distributed with v2.12 of IRAF. Bias
subtraction and flat fielding were the main processing steps. While
only single exposures were obtained for most fields, effort was made
to identify saturated pixels and defects caused by cosmic rays,
scattered light, satellite trails, etc., and to map the locations of
affected pixels in a bad pixel mask for each image. These masks were
used during the generation of the catalogues, described below, to
flag objects whose properties could not be well measured.

\subsubsection{Astrometry}

In order to generate relatively accurate positions for the galaxies
whose properties are presented in the electronic Tables~3-18, we
needed to determine for each of the images an accurate world
coordinate system.  This was generated using software in IRAF (tasks
ccsetwcs and msccmatch in the mscred package; as described by
\citet{2002adaa.conf..309V}) and the USNO A2.0 catalogue of the
positions of objects in each of the fields
\citep{1998yCat.1252....0M}.  A limited magnitude range was used of
USNOA2.0 reference stars in order to avoid the documented relative
offset in astrometry for this catalog as a function of magnitude.
Using the IRAF task msccmatch (in v4.7 of the mscred package), a
fourth order mapping between the images in x,y pixel coordinates to
RA and DEC of the reference stars was determined.  The RMS residuals
of the fit were 0.3 to 0.4$''$ for each of the images. At the time
these solutions were determined, the USNO A2.0 was the best
catalogue available. Although catalogues with improved positions are
now available (e.g. USNO-B1.0, \citet{2003AJ....125..984M} or GSC2.2
\footnote{The Guide Star Catalogue-II is a joint project of the
Space Telescope Science Institute and the Osservatorio Astronomico
di Torino. Space Telescope Science Institute is operated by the
Association of Universities for Research in Astronomy, for the
National Aeronautics and Space Administration under contract
NAS5-26555.  The participation of the Osservatorio Astronomico di
Torino is supported by the Italian Council for Research in
Astronomy. Additional support is provided by European Southern
Observatory, Space Telescope European Coordinating Facility, the
International GEMINI project and the European Space Agency
Astrophysics Division.}), a trial new solution using the USNO-B1.0
did not result in a significantly improved solution, so we have
retained our original fits.  The residuals to our solutions were
largest around the border of the images and we suspect that the
uncertainties in the positions of objects around the edge of each
field should probably include an additional 0.1 to 0.2$''$ of
systematic error. Our measured positions of objects given in this
paper are in the ICRF and J2000.

\subsubsection{Photometry and Catalogue Generation}

We did not obtain multiple standard star observations nor observe
in a range of colours that would allow us to determine our own
extinction solutions and colour terms.  This fundamentally limits
the accuracy we can obtain in our photometry. However, our needs
in the current study only require photometry accurate to a few
tenths of a magnitude. Using a representative Mauna Kea extinction
curve determined from photometric observations during other runs
and our observations of the stars in NGC~7790, which have been
calibrated in the $R-$band by \citet{1992PASP..104..553O}, we were
able to determine zero points for the 10 fields imaged during our
photometric nights.

There were five fields observed only on non-photometric nights. We
have determined an approximate zero-point for these images as
follows. First, for the 10 fields we calibrated with our measurements
of NGC~7790, we compared our measured magnitudes to the red magnitudes
of the GSC2.2 for objects in common. The pass-bands, of course, are
not the same, and we would not expect the magnitudes to be identical,
but we were gratified that all of the fields observed under what we
believe to have been photometric conditions yielded similar offsets
(mean offset of 0.1 of a magnitude) and RMS residuals (0.15 to
0.25). We then assigned to each of the five fields (all from the 1997
run) observed on non-photometric nights a zero-point based on the
calibration determined from the photometric night of that run.  We
then compared these five fields to the GSC2.0. Two of the fields,
3C~57 and PKS~2340$-$036, had mean differences in the same range as
those fields observed under photometric conditions. Three of the
fields; NGC 2841 UB3, 4C~41.21, and PKS 2344$+$09; had significantly
larger initial offsets (0.43, 0.47, and 0.62 magnitudes). The RMS
residuals, however, were similar to those of all the other fields.  We
therefore adjusted the zero-points for these fields so that their mean
offset when compared to the GSC2.0 stars would also be 0.1 of a
magnitude. This adjustment is consistent with what is evident from
inspecting the images (and number counts), namely that these exposures
taken with some clouds did not reach the same depth (in the same
exposure time) as our other exposures.  A systematic uncertainty of
0.2 mags should probably be assigned to our reported magnitudes for
observed galaxies, but we feel we have successfully placed our
observations on a common relative photometric scale.

Using our magnitude zero-points for each image and measurements of
the variance in the sky brightness of each image, we have confirmed
that the five sigma detection thresholds for point sources in all of
our images is a magnitude of 24.9 or greater, well below our
selection of targets for spectroscopic follow-up.

For each field the object catalogues were generated using SExtractor
2.2.2 \citep{1996A&AS..117..393B} run with the minimum detection
area, Gaussian convolution filter, and signal above sky threshold
optimized to detect and measure all of the objects for which
spectroscopic data was obtained.  We measured eight fixed aperture
magnitudes (starting at 2$''$ and increasing to 10$''$ in diameter)
and the SExtractor MAG\_BEST, which is similar to the Kron total
magnitude \citep{1980ApJS...43..305K}. Results for our
spectroscopically observed galaxies are included in the electronic
Tables~3-18.

As previously discussed, for some of our fields scattered light
affected our images. Even without these problems, any catalog will
suffer increasing incompleteness for galaxies of fainter total
magnitude and/or surface brightness.  As a test of the completeness
of our imaging catalogs, artificial galaxies were inserted into each
image (using the IRAF artdata task) and SExtractor used to attempt
to recover the objects. While not a perfect measure of the
robustness of our catalogs, the 50\% completeness limit of our
catalogs for each quasar field is typically 24.4 ($R-$band) and
greater than 24.0 for all fields. The fields most seriously affected
by the arcs of scattered light could be missing even bright galaxies
at the 5\% level, but visual inspection indicates that the fraction
is likely to be much smaller.

\subsubsection{Bounded Restframe $B$-band Magnitude} \label{subsubsec-absmag}

We would like to have well measured luminosities for each of the
galaxies expressed in a common rest-frame band.  Unfortunately, we
do not have well measured colours or spectrophotometry that would
allow an accurate determination of the over-all spectral energy
distributions (SEDs) of the galaxies in our sample.  This prevents
the assignment of the proper $K-$correction for each galaxy in our
sample.

We can, however, constrain the range of rest frame $B-$band
luminosities possible for each galaxy using our $R-$band
measurements, the redshifts determined from our spectroscopy, and
the galaxy SED templates and $K-$corrections of
\citet{1980ApJS...43..393C}.  For each galaxy we determined the
range of possible $K-$corrections and intrinsic $B-R$ colours and
tabulated the resultant maximum and minimum rest-frame $B-$band
magnitudes.  These are listed for each galaxy in the electronic
Tables~3-18. We assumed no evolution or intrinsic extinction.  While
more involved procedures might, with other data sets, provide
probability distributions for the rest $B-$band luminosities of each
galaxy, the procedure described above allows us to make the gross
comparisons between sub-samples that are reasonable given the other
uncertainties in our measurements.  In other words, those galaxies
that might be the intrinsically more luminous are separable from
those that could not. The observed redshifts and $R-$band magnitudes
are the dominant constraints.

\subsection{MOS Spectroscopy} \label{subsec-spect}

\subsubsection{Observing}

The majority of the galaxy redshifts described in this paper were
obtained using the CFHT Multi-Object Spectrograph (MOS),
\citep{1994A&A...282..325L}, during observing runs in July 1995 and
November 1997. As described above, imaging catalogs were prepared in
real time, from which masks were designed. For July 1995 the Loral 3
CCD was used along with the O300 grism, yielding roughly 3.5\AA\ per
pixel. No wavelength blocking filter was used, yielding a wavelength
coverage that varied depending on the location of the slit within
the field of view, but reasonable signal-to-noise (S/N) was
generally obtained for wavelengths between 5000~\AA\ and 9000~\AA.
For the November 1997 run, the STIS 2 CCD was used with the same
grating, yielding a dispersion of 5.2\AA\ per pixel. for both runs,
masks were designed with a 1.5\arcsec\ slit width, corresponding to
4.8 pixels in July 1995 and 3.4 pixels in November 1997. Some of the
observations were taken through cloud. A flux standard star was
observed in order to allow an approximate flux calibration of the
data and to allow us to remove the instrumental signature, but
because of the above mentioned cloud, the fluxes recorded could be
low.

Spectroscopic multi-slit masks were designed using the galaxy
catalogs generated during the observing run. Slits were assigned to
galaxies with a magnitude ranking. Slits had a default length of
10\arcsec\ and width of 1.5\arcsec . After this, the assignment
algorithm checked whether it was possible to add further objects by
sliding slits sideways by a limited amount. Finally all slits were
expanded in length to fill the array and allow better sky
subtraction for some objects. Depending on the galaxy surface
density and distribution in the field between 40 and 47 galaxies had
slits assigned to them per mask in this manner. Between one and
three masks per LOS were taken.

Because of the varying conditions mentioned above, the redshift
success rate for objects which were assigned slits varies from mask
to mask. The average success rate was 0.5, with a nadir of 0.07 for
a mask taken through thick cloud, and a pinnacle of 0.89 for one
particularly good mask.

It was never our intent to obtain a sample complete to any given
magnitude, but merely to obtain a statistically useful sample of the
galaxies around the LOS. By comparing with the photometric catalog,
we can reconstruct our completeness for each field as a function of
magnitude, as will be discussed below.

\subsubsection{Reduction}

The spectroscopic data reduction using IRAF followed the standard
procedure for MOS spectroscopy used for example in the CNOC2 project
\citep{2000ApJS..129..475Y}. The data were bias subtracted and
trimmed. If necessary, bad columns were interpolated over. Repeat
exposures of the same mask were combined and cosmic rays removed.
Because the spectral extraction involves summing over several
columns, flat fielding to correct for pixel-to-pixel variations in
CCD sensitivity was not necessary (and would have added noise).
Individual apertures were then summed to generate 1D spectra,
simultaneously subtracting the sky using adjacent regions along the
slit. These spectra were wavelength calibrated using arc spectra
obtained at the same position on the sky (to minimize the effects of
flexure), and the spectra were approximately flux calibrated using
observations of a standard star taken through the same slit width.

\subsubsection{Redshift Determination}

The driving goal for the redshift measurements was to obtain
redshifts that could all be placed in the same reference frame, in
our case heliocentric. To do this, all the galaxy redshifts
determined in different ways were tied together in the same
reference frame, and then this frame was tied to that of the
absorption line lists, which were adjusted to the heliocentric
velocity frame as described by  Jannuzi et al. 1998.

All galaxies were cross correlated with two absorption line
templates using the IRAF XCSAO task. One template was the
`fabtemp97' one which comes with the XCSAO package, and the other
was a high-quality wide-wavelength coverage spectrum of the
early-type galaxy NGC 4889, kindly provided by B. Oke (private
communication, template available on request from S. Morris). As
usual, the cross correlation process involved continuum fitting and
also filtering of high and low Fourier components. Redshifts for
emission line objects were obtained more interactively using the
IRAF RVIDLINES task inside the RV package. The results of these
routines were then checked by eye, and each measurement flagged as
successful or failed. This somewhat subjective procedure was
necessary in order to deal with the combined effects of zero order
contamination, poor sky subtraction, residual cosmic rays and other
reduction problems. In practice, for the cross correlation
measurements, there was a fairly clear threshold in R-value reported
by the routine above which  between measurement were judged
successful. Despite all of the above, it is still of course possible
for some objects in our redshift catalog to have `catastrophic'
redshift errors (i.e. redshifts based on spurious features of
mis-identified features). We believe that the above procedure should
have minimized the number of such errors, and would be surprised if
there were more than a handful these in the redshift catalog, but it
should be noted that the redshift errors do not take account of this
possibility.

Both the above routines automatically convert redshifts to a
heliocentric frame. Nevertheless, as is well known, this does not
mean that there are no systematic shifts in velocity introduced by
the different processes, and so objects with reliable redshifts
determined from 2 or more of the above approaches were used to
measure and remove any such shifts. (Shifts of -49 \kms\ and +72
\kms\ were measured for the NGC 4889 and emission line redshifts
relative to the fabtemp97 measurements).

At this stage we are left with 641 galaxy redshifts in our catalog
with their associated redshift errors. The success rate of
obtaining redshifts from a given number of slits on a mask varied
considerably depending on conditions and the magnitude range of
objects on the masks, but for typical masks and conditions was
around 60-70\%.

As described above, the photometry performed on the mountain for
mask design was repeated after the observing run. The resulting
photometric catalog was then re-matched with the redshift catalog,
resulting in our dropping two of our measured objects because of
large uncertainties in their positions and magnitudes. The CFHT
MOS data therefore provided 639 new redshifts of galaxies in the
targeted LOS.

The objects with identified redshifts are listed in a set of Tables
(3-18) available in the electronic version of the paper. The columns
in the associated electronic table are as follows:

\begin{enumerate}
\item RA: The J2000 Right ascension of the galaxy. The relative
positions for galaxies in a single quasar field should be good to a
few tenths of an arcsecond.

\item DEC: The J2000 Declination of the galaxy. The relative positions
for galaxies in a single quasar field should be good to a few tenths
of an arcsecond.

\item z: The redshift determined from our MOS spectrum of the galaxy.

\item z$_{\rm err}$: this is the error reported by the cross
correlation routine, and as such should be considered a lower limit
on the redshift error.

\item R-mag: R-band apparent magnitude (MAG\_BEST from Sextractor).
As discussed above, these are only accurate to $\pm$0.2 magnitudes.

\item B$_{\rm max}$: estimate of the maximum rest frame B band
luminosity as discussed in \S~\ref{subsubsec-absmag}

\item B$_{\rm min}$: estimate of the maximum rest frame B band
luminosity as discussed in \S~\ref{subsubsec-absmag}
\end{enumerate}

\begin{table*}
\begin{minipage}{140mm}
\caption{Sample Table - CFHT MOS Sample Galaxy Properties - NAB
0024+22} \label{tab-gal_NAB0024+22}
\begin{tabular}{@{}ccccccc}
\hline
R.A.(J2000)& Decl.(J2000) & z     & z$_{\rm err}$  & R-mag    & B$_{\rm max}$   & B$_{\rm min}$ \\
\hline
0:26:55.8 & 22:41:02.3 & 0.1587 & 0.0003 & 18.1 & -20.4 & -19.6 \\
0:26:56.7 & 22:40:09.3 & 0.3100 & 0.0006 & 20.2 & -19.9 & -19.3 \\
0:26:57.8 & 22:41:30.6 & 0.3469 & 0.0007 & 21.0 & -19.3 & -18.9 \\
... & ... & ... & ... & ... & ... & ... \\
\hline
\end{tabular}
\end{minipage}
\end{table*}

\setcounter{table}{18}

\subsection{Additional Galaxy Redshifts} \label{subsec-add}

We also searched the literature to find additional galaxies with
measured redshifts around the sample QSO LOS. As the typical
redshift range for the SDSS and 2dF galaxy redshift samples are not
well matched to absorption line data used in this paper, we have not
chosen to add in galaxies from those surveys at this stage.

Our main method for finding additional redshifts was through the NED
database\footnote{This research has made use of the NASA/IPAC
Extragalactic Database (NED) which is operated by the Jet Propulsion
Laboratory, California Institute of Technology, under contract with
the National Aeronautics and Space Administration.}. We list the
results of this search below. Since the analysis in this paper is
statistical in nature, we have chosen only to add redshifts from
other authors to our sample when they make a significant difference
to the number of redshifts known for a given LOS.

\subsubsection{PKS 0122$-$00}

Galaxies in this LOS were observed by both
\citet{1997MNRAS.284..599B}, and \citet{2001ApJ...559..654C}, but
as only two additional galaxy redshifts would have been added to
the sample, for simplicity these were not included. For the three
galaxies in common between our sample and that in
\citet{2001ApJ...559..654C} a mean velocity difference of 30 \kms\
is measured, with an RMS of 450 \kms\ .

\subsubsection{PKS 2145$+$06}

A galaxy that seems to be associated with a MgII absorber at
z$\sim$0.79 in the LOS has been studied in a number of papers,
e.g. \citep{2000ApJS..130...91C, 2000ApJ...543..577C}. This galaxy
has not been added to the sample.

\subsubsection{PG 2302+029}

This LOS was surveyed by \citet{2001ApJ...547...39B}, obtaining
spectra for 24 objects. In this same field, we measured redshifts
for 42 objects. We had 6 galaxies in common. (For which we measure a
reassuringly small mean velocity shift of -65 \kms\ with RMS 227
\kms\ , in line with our error estimates). This then gives us 18
additional redshifts which we add to our sample. These objects are
included in (the electronic) Table~16. On a less positive note, it
is interesting that 2 groups designing masks for the same LOS could
end up with so little overlap in their samples.

\subsubsection{Q0107$-$025A,B}

The region around the LOS to the QSO pair Q0107-025A,B was also
observed with the COSMIC spectrograph on the Palomar 5m telescope
(http://www.astro.caltech.edu/palomar/200inch/cosmic/) by Weymann et
al. (private communication). The measured redshifts for one of the
masks observed was kindly made available to us by M. Rauche. This
yielded 28 additional objects with redshifts. In this case there was
no overlap by design, and so we are assuming that any systematic
shifts between the two redshift sets is small. These objects are
included in (the electronic) Table~4.

\clearpage

\begin{figure}
\includegraphics[width=84mm]{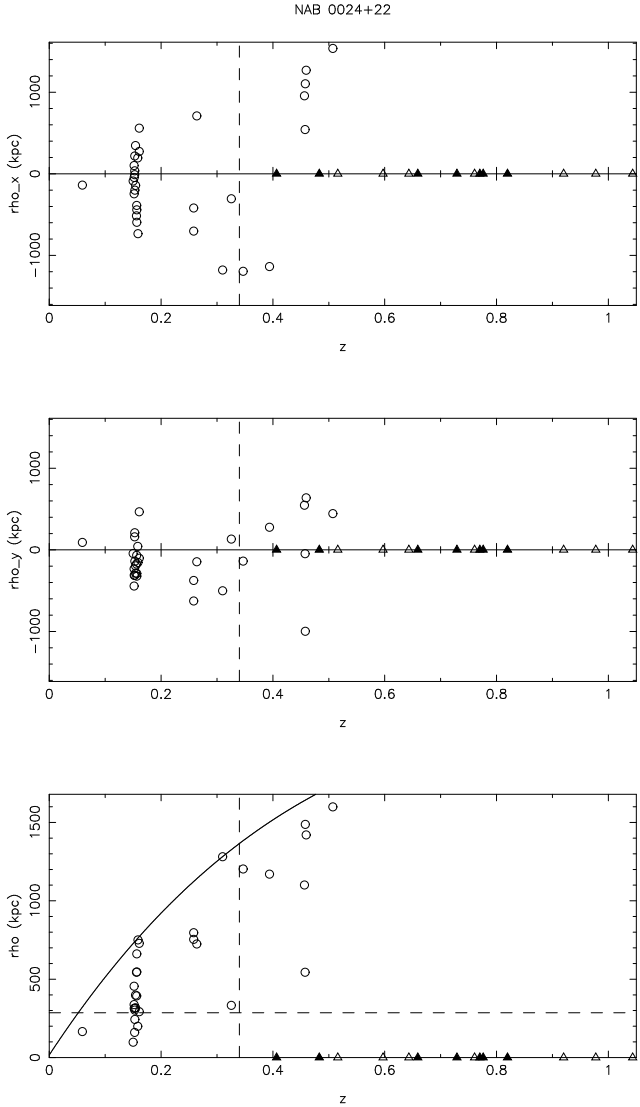}
\caption{Pie diagram showing absorbers and galaxies in the LOS of
NAB~0024$+$22. The top and middle panels show projections of the
galaxy distribution in RA and Dec, while the bottom panel shows the
galaxy distribution in impact parameter as a function of redshift.
Open circles represent galaxies, open triangles are absorbers
detected in \lya, filled triangles are absorbers seen in CIV. A star
(if plotted) shows the location of the QSO. In all three panels, the
vertical dashed line at z=0.34 shows the approximate location of the
lowest redshift \lya\ detectable with the HST FOS G190H grating. In
the bottom panel, the horizontal dashed line shows an impact
parameter of 280 kpc in our adopted cosmology (see
\S~\ref{subsec-pairid}), the curved solid line shows the maximum
impact parameter observable with the CFHT MOS FOV, while the
vertical dotted lines join absorber/galaxy `pairs' conservatively
identified by their being within 1000 \kms\ of each other in
redshift.}
  \label{fig-n0024pie}
\end{figure}

\begin{figure}
\includegraphics[width=84mm]{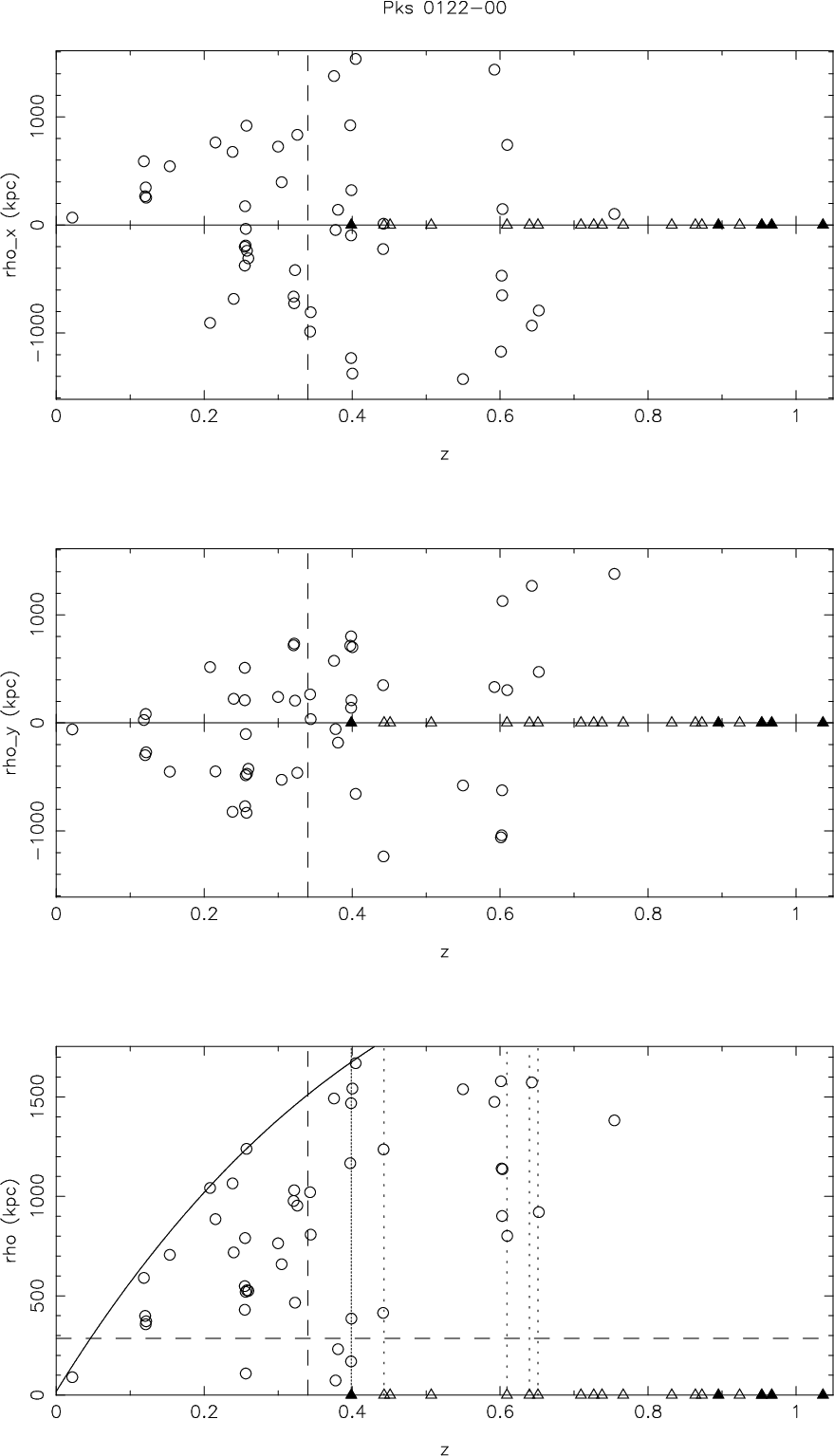}
 \caption{Pie diagram showing absorbers and galaxies in the LOS of PKS~0122$-$00.
Description of figure as per figure~\ref{fig-n0024pie}.}
  \label{fig-p0122pie}
\end{figure}

\begin{figure}
\includegraphics[width=84mm]{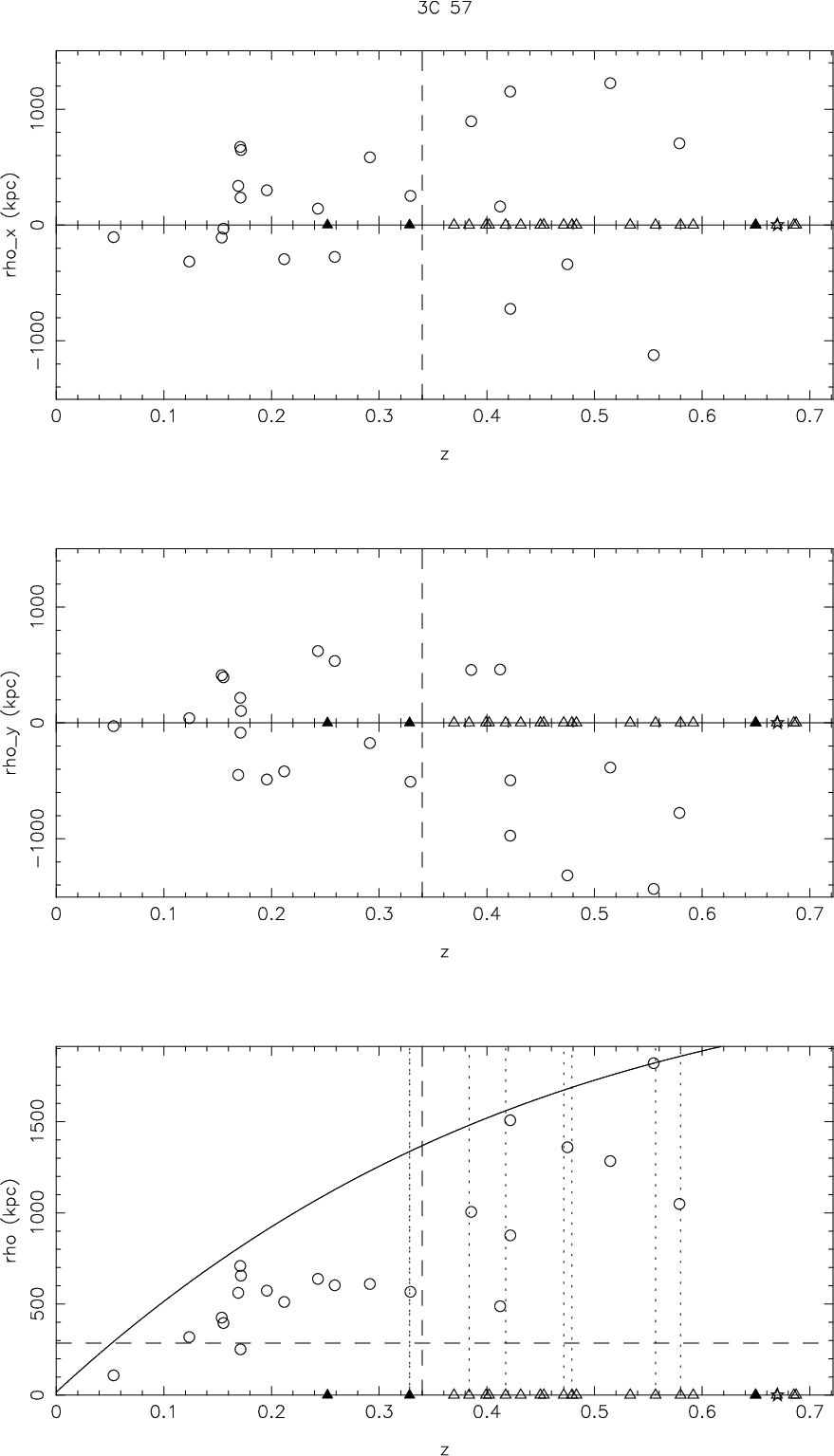}
 \caption{Pie diagram showing absorbers and galaxies in the LOS of 3C~57.
Description of figure as per figure~\ref{fig-n0024pie}.}
  \label{fig-3c057pie}
\end{figure}

\begin{figure}
\includegraphics[width=84mm]{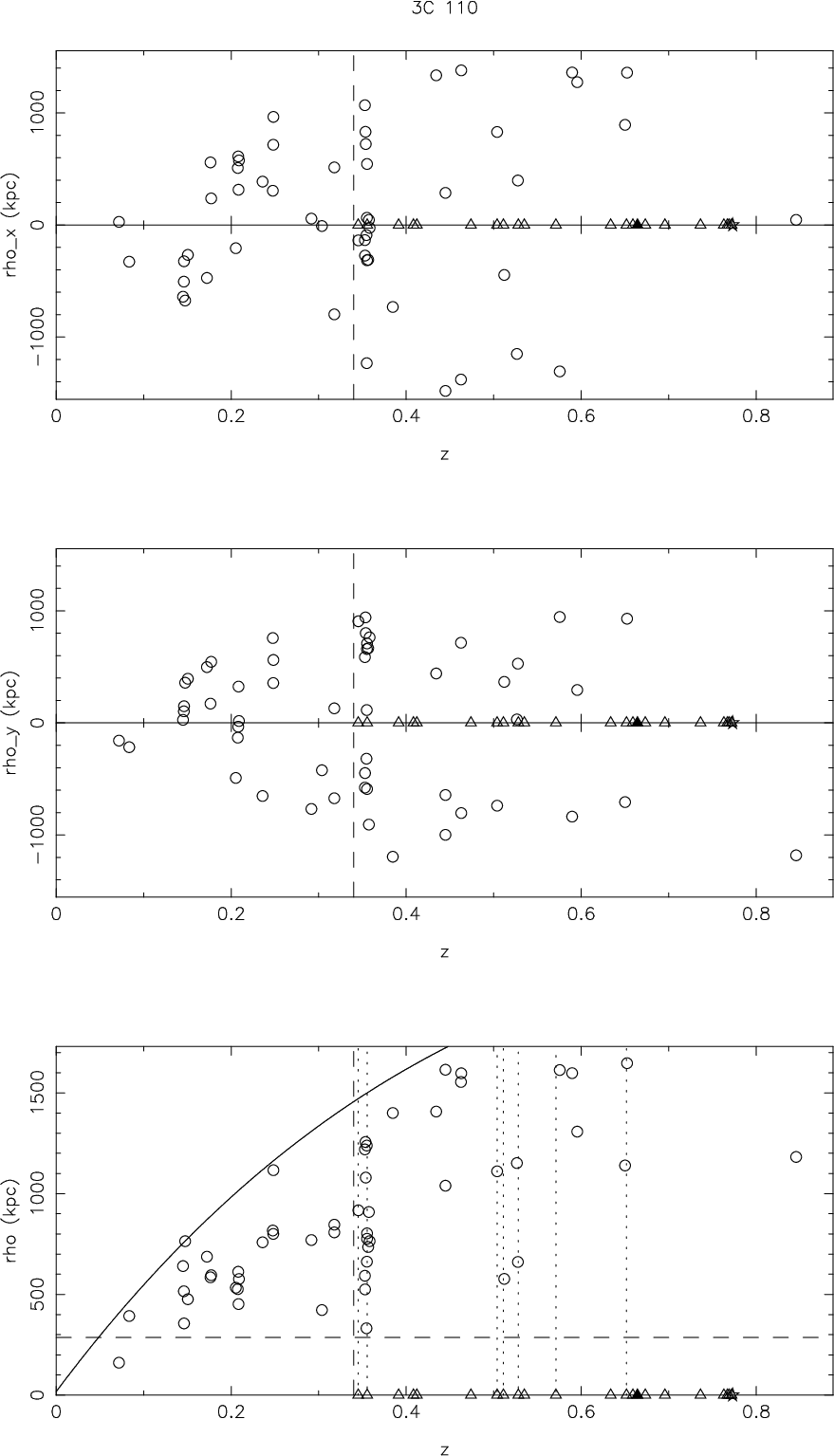}
 \caption{Pie diagram showing absorbers and galaxies in the LOS of 3C~110.
Description of figure as per figure~\ref{fig-n0024pie}.}
  \label{fig-3c110pie}
\end{figure}

\begin{figure}
\includegraphics[width=84mm]{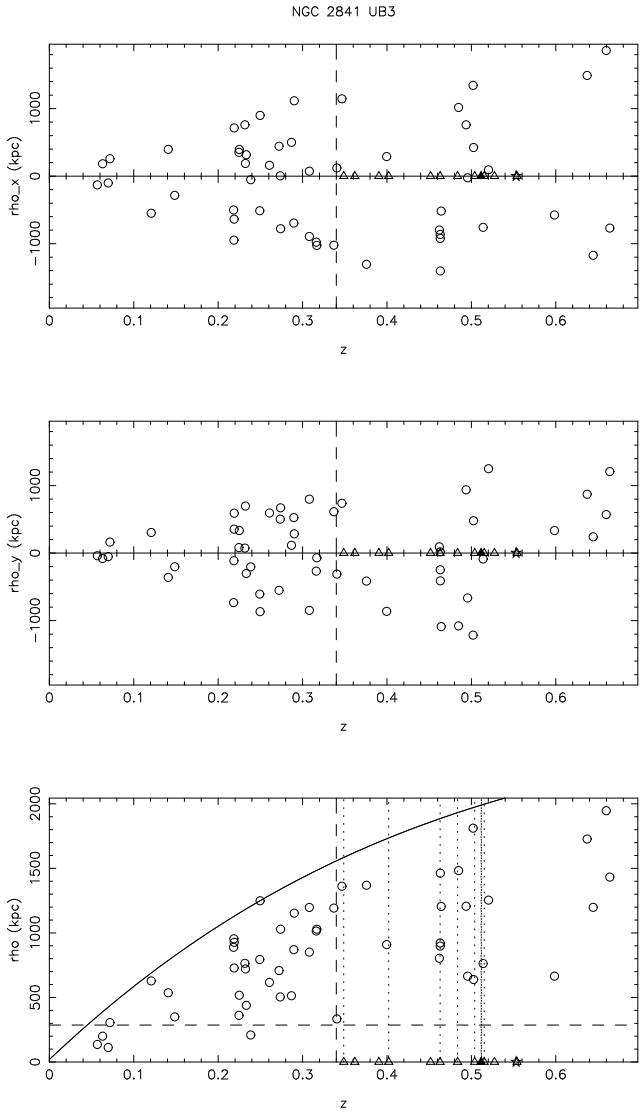}
 \caption{Pie diagram showing absorbers and galaxies in the LOS of NGC~2841~UB3.
Description of figure as per figure~\ref{fig-n0024pie}.}
  \label{fig-n2841pie}
\end{figure}

\begin{figure}
\includegraphics[width=84mm]{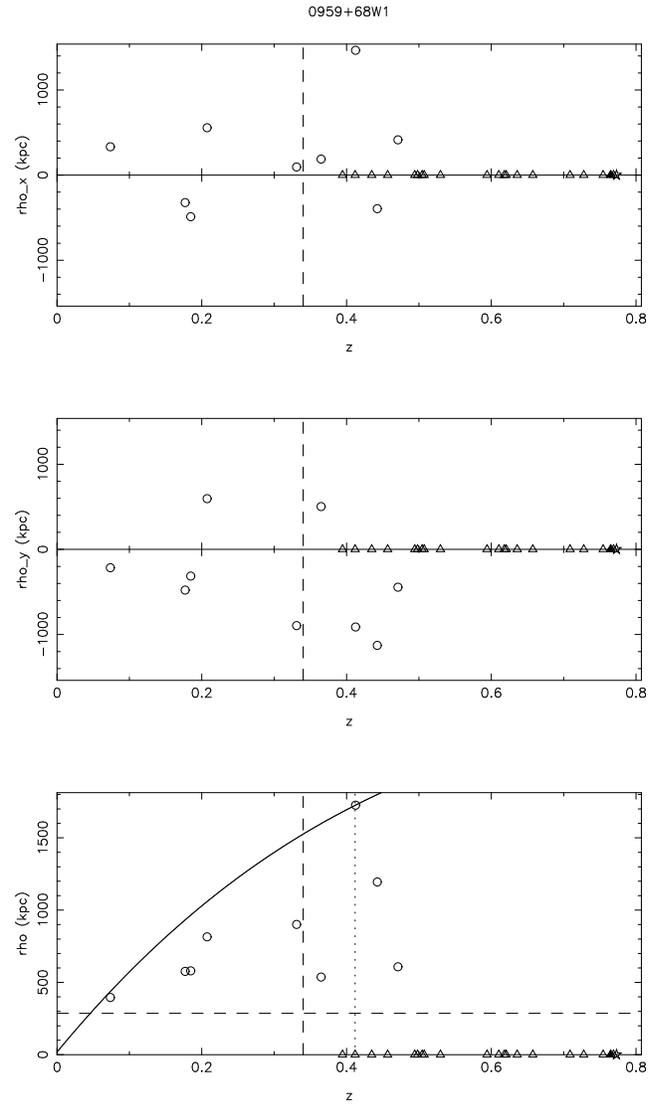}
 \caption{Pie diagram showing absorbers and galaxies in the LOS of 0959$+$68W1.
Description of figure as per figure~\ref{fig-n0024pie}.}
  \label{fig-q0959pie}
\end{figure}

\begin{figure}
\includegraphics[width=84mm]{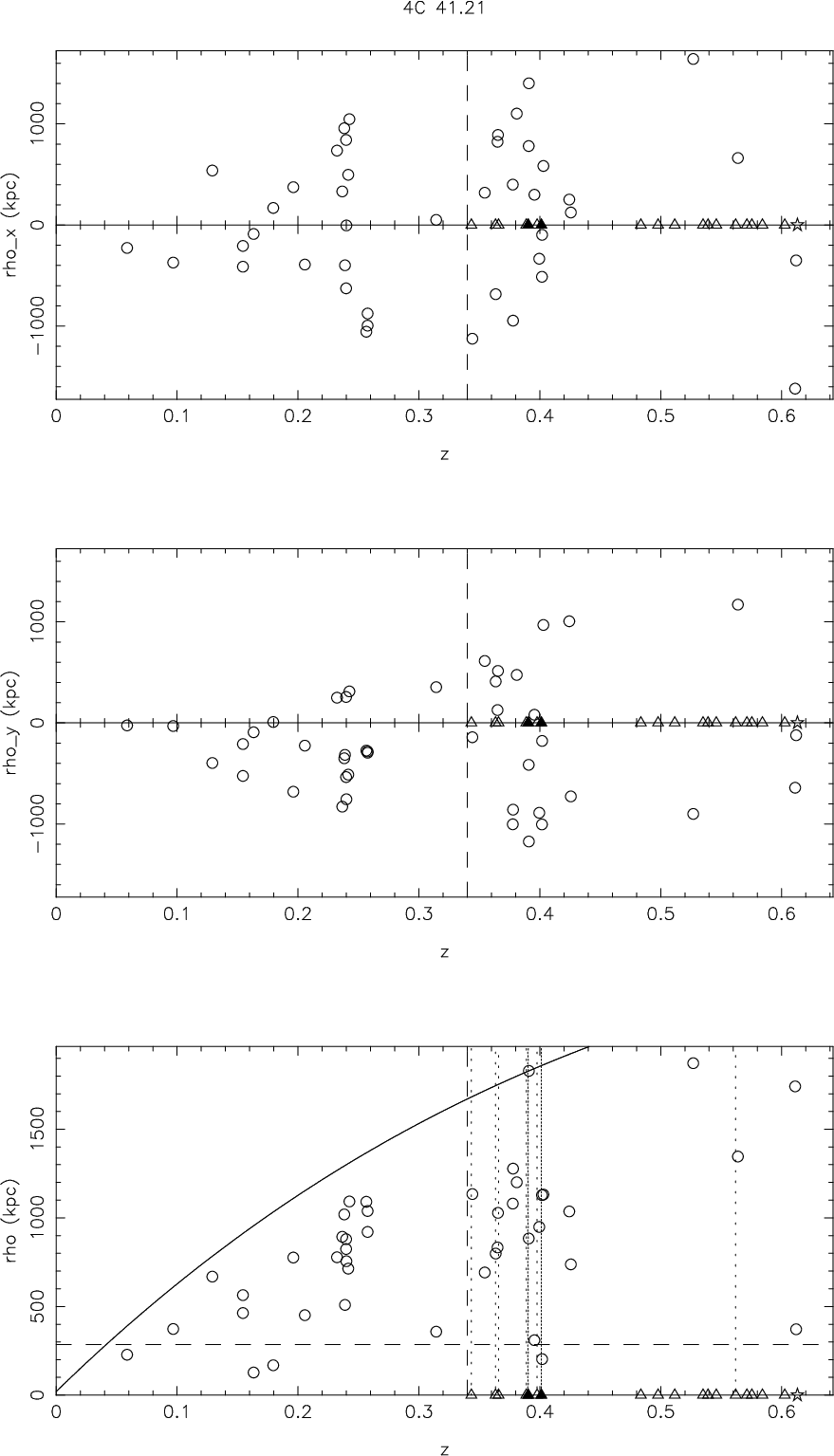}
 \caption{Pie diagram showing absorbers and galaxies in the LOS of 4C~41.21.
Description of figure as per figure~\ref{fig-n0024pie}.}
  \label{fig-4c041pie}
\end{figure}

\begin{figure}
\includegraphics[width=84mm]{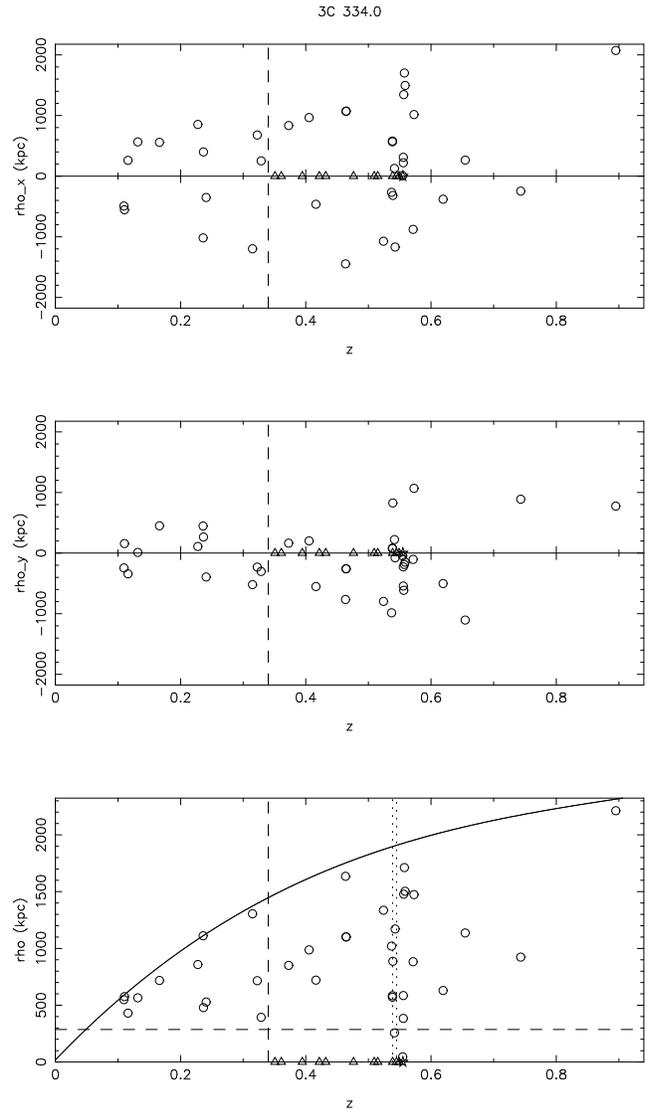}
 \caption{Pie diagram showing absorbers and galaxies in the LOS of 3C~334.
Description of figure as per figure~\ref{fig-n0024pie}.}
  \label{fig-3c334pie}
\end{figure}

\begin{figure}
\includegraphics[width=84mm]{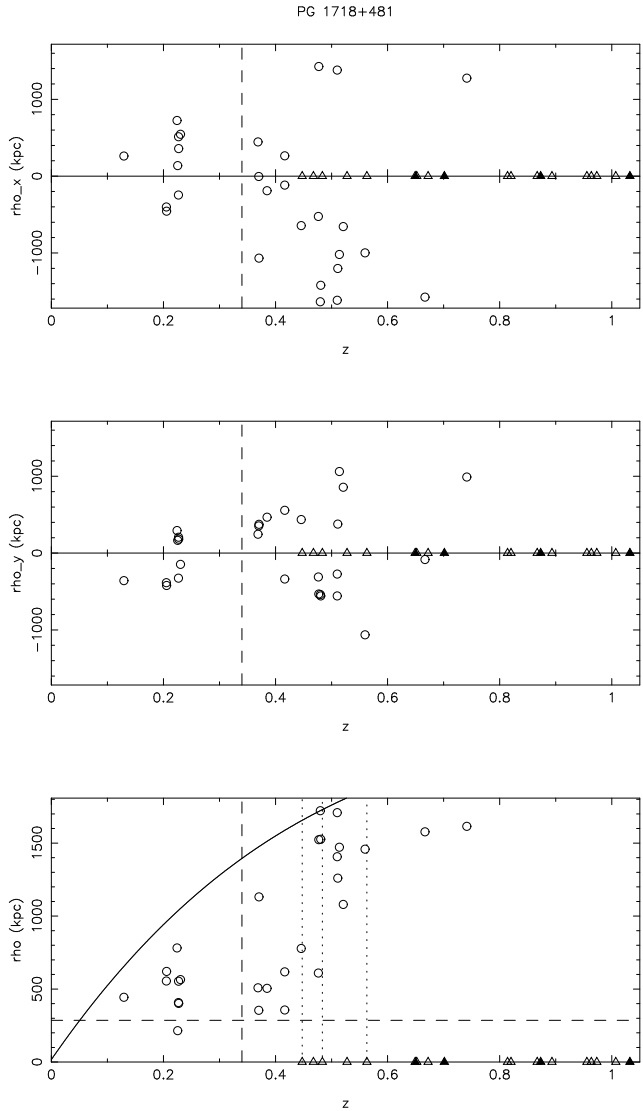}
 \caption{Pie diagram showing absorbers and galaxies in the LOS of PKS~1718$+$481.
Description of figure as per figure~\ref{fig-n0024pie}.}
  \label{fig-p1718pie}
\end{figure}

\begin{figure}
\includegraphics[width=84mm]{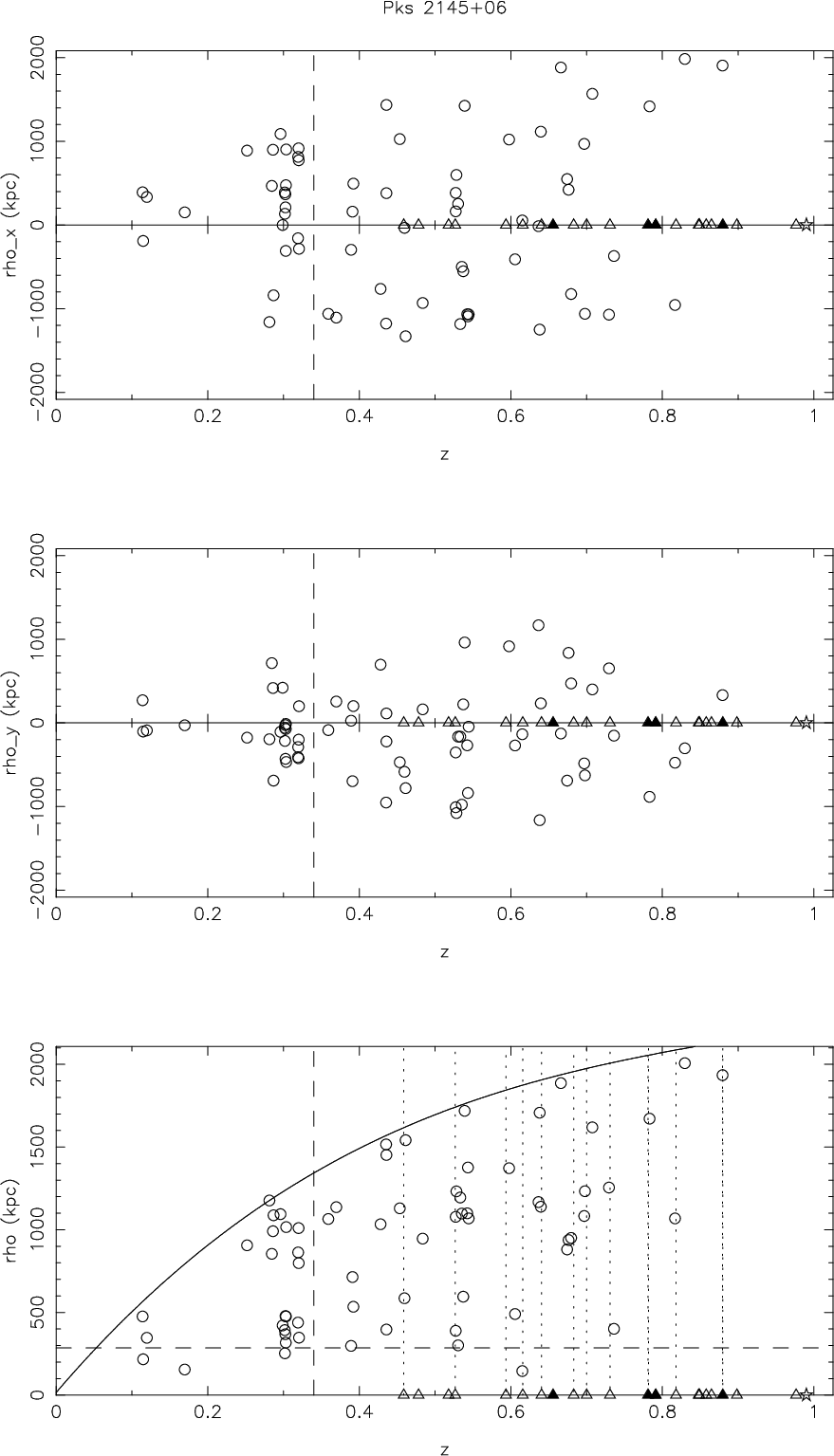}
 \caption{Pie diagram showing absorbers and galaxies in the LOS of PKS~2145$+$06.
Description of figure as per figure~\ref{fig-n0024pie}.}
  \label{fig-p2145pie}
\end{figure}

\begin{figure}
\includegraphics[width=84mm]{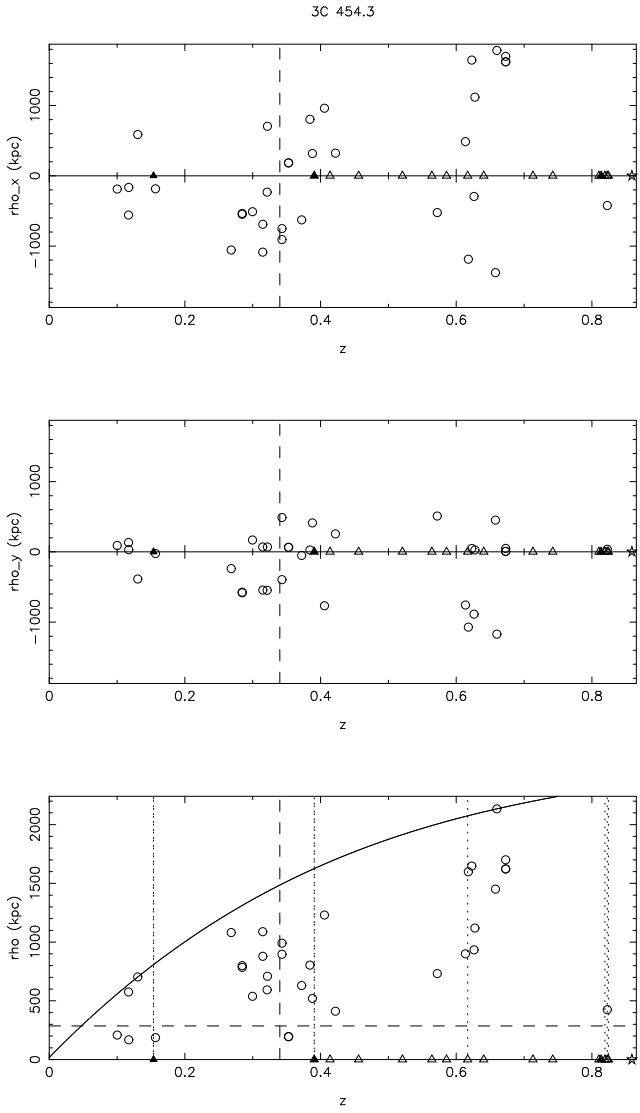}
 \caption{Pie diagram showing absorbers and galaxies in the LOS of 3C~454.
Description of figure as per figure~\ref{fig-n0024pie}.}
  \label{fig-3c454pie}
\end{figure}

\begin{figure}
\includegraphics[width=84mm]{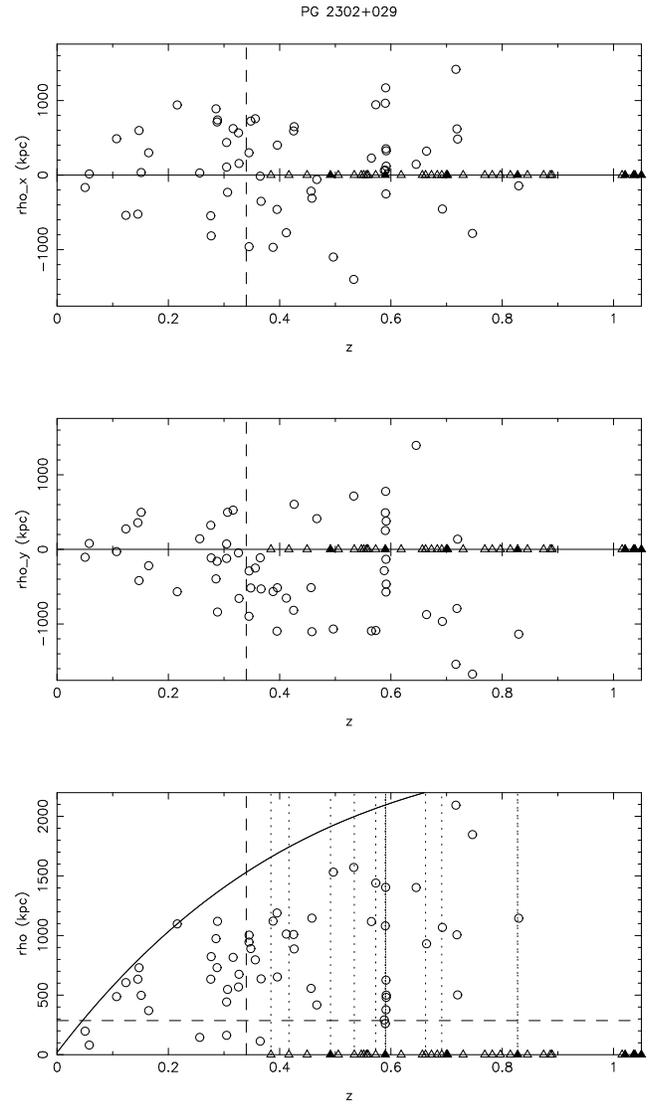}
 \caption{Pie diagram showing absorbers and galaxies in the LOS of PG~2302$+$029
 (includes the additional objects from Bowen et al.~(2001)).
Description of figure as per figure~\ref{fig-n0024pie}.}
  \label{fig-p2302pie}
\end{figure}

\begin{figure}
\includegraphics[width=84mm]{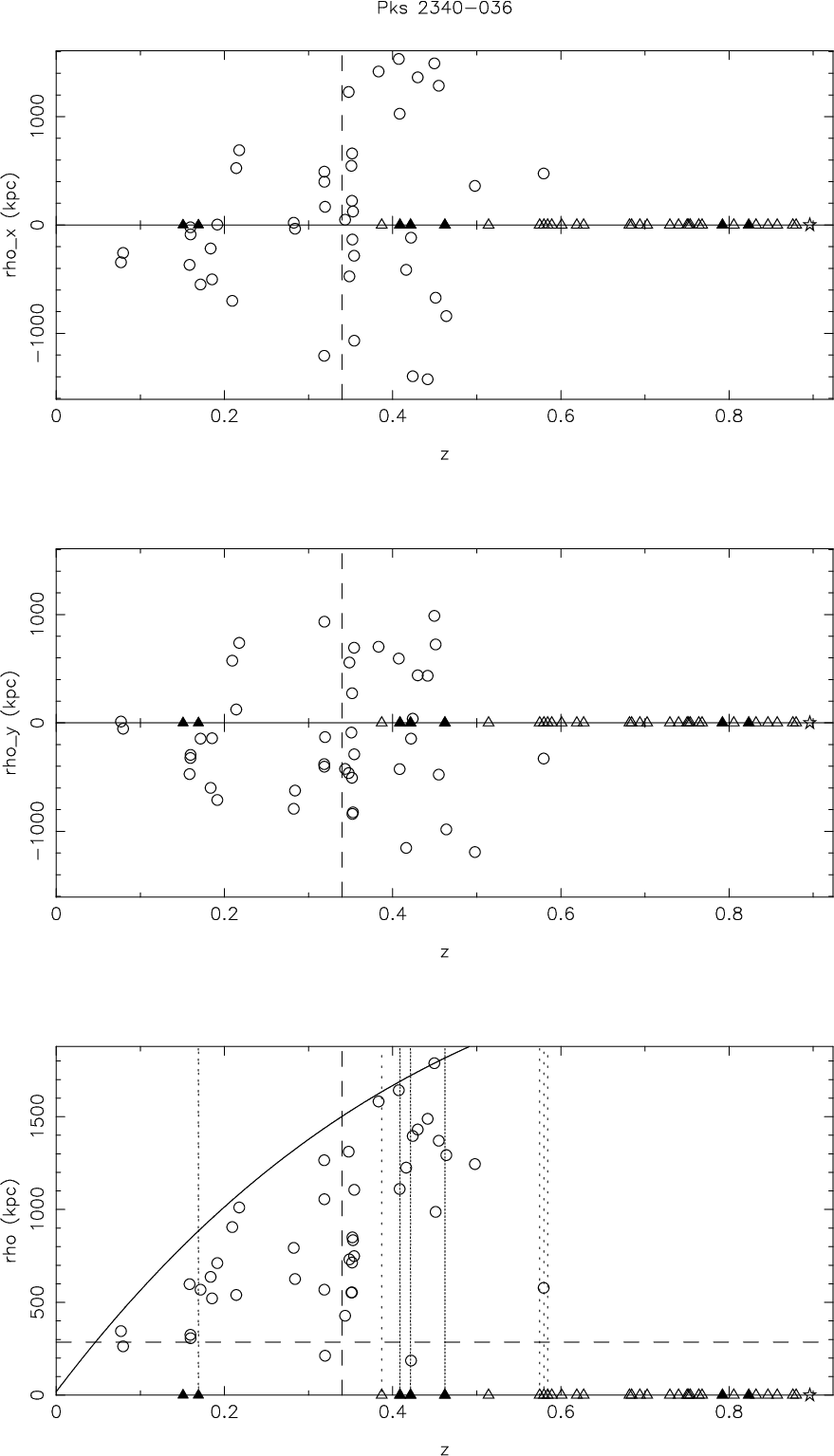}
 \caption{Pie diagram showing absorbers and galaxies in the LOS of PKS~2340$-$036.
Description of figure as per figure~\ref{fig-n0024pie}.}
  \label{fig-p2340pie}
\end{figure}

\begin{figure}
\includegraphics[width=84mm]{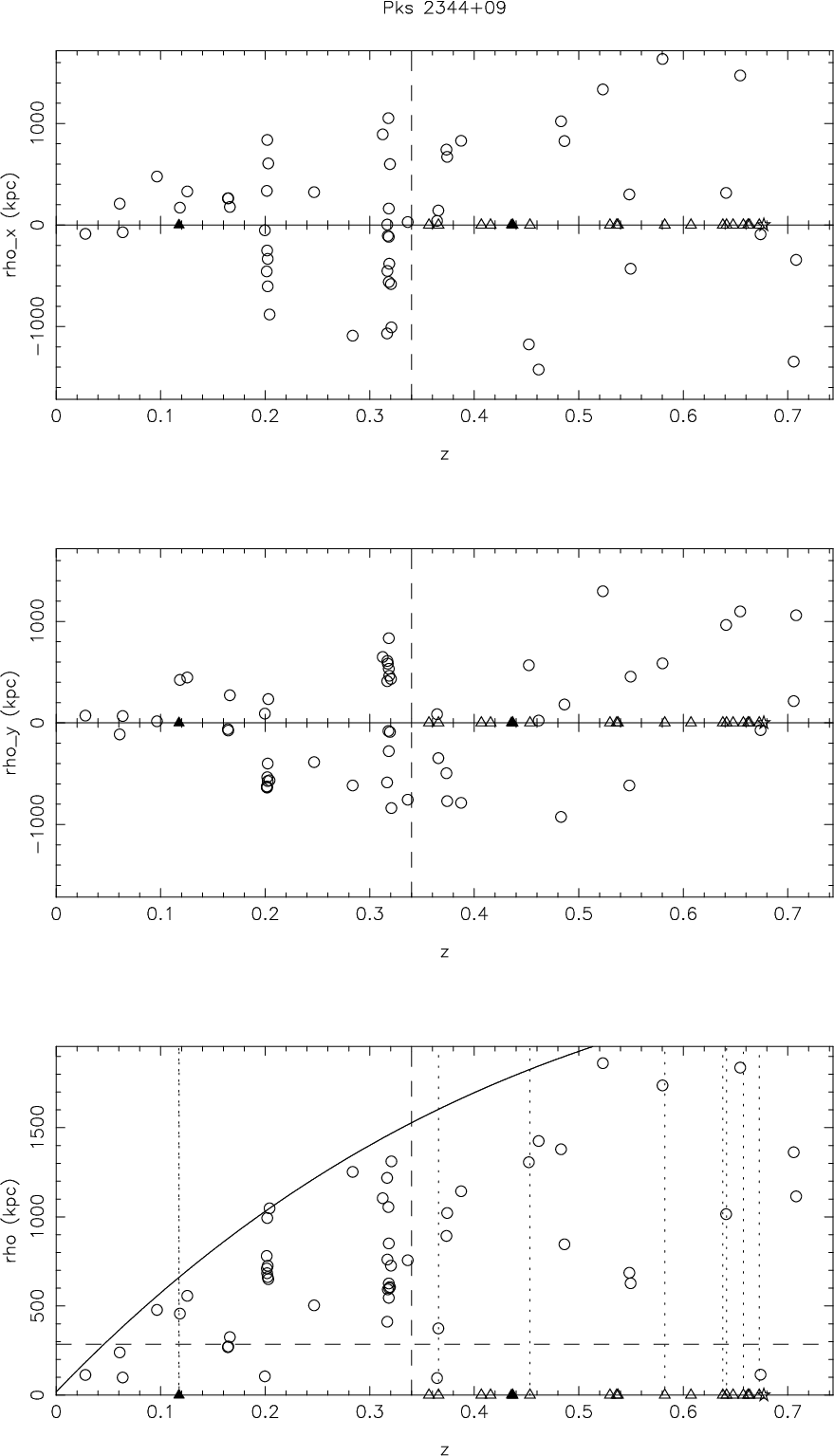}
 \caption{Pie diagram showing absorbers and galaxies in the LOS of PKS~2344$+$09.
Description of figure as per figure~\ref{fig-n0024pie}.}
  \label{fig-p2344pie}
\end{figure}

\begin{figure}
\includegraphics[width=84mm]{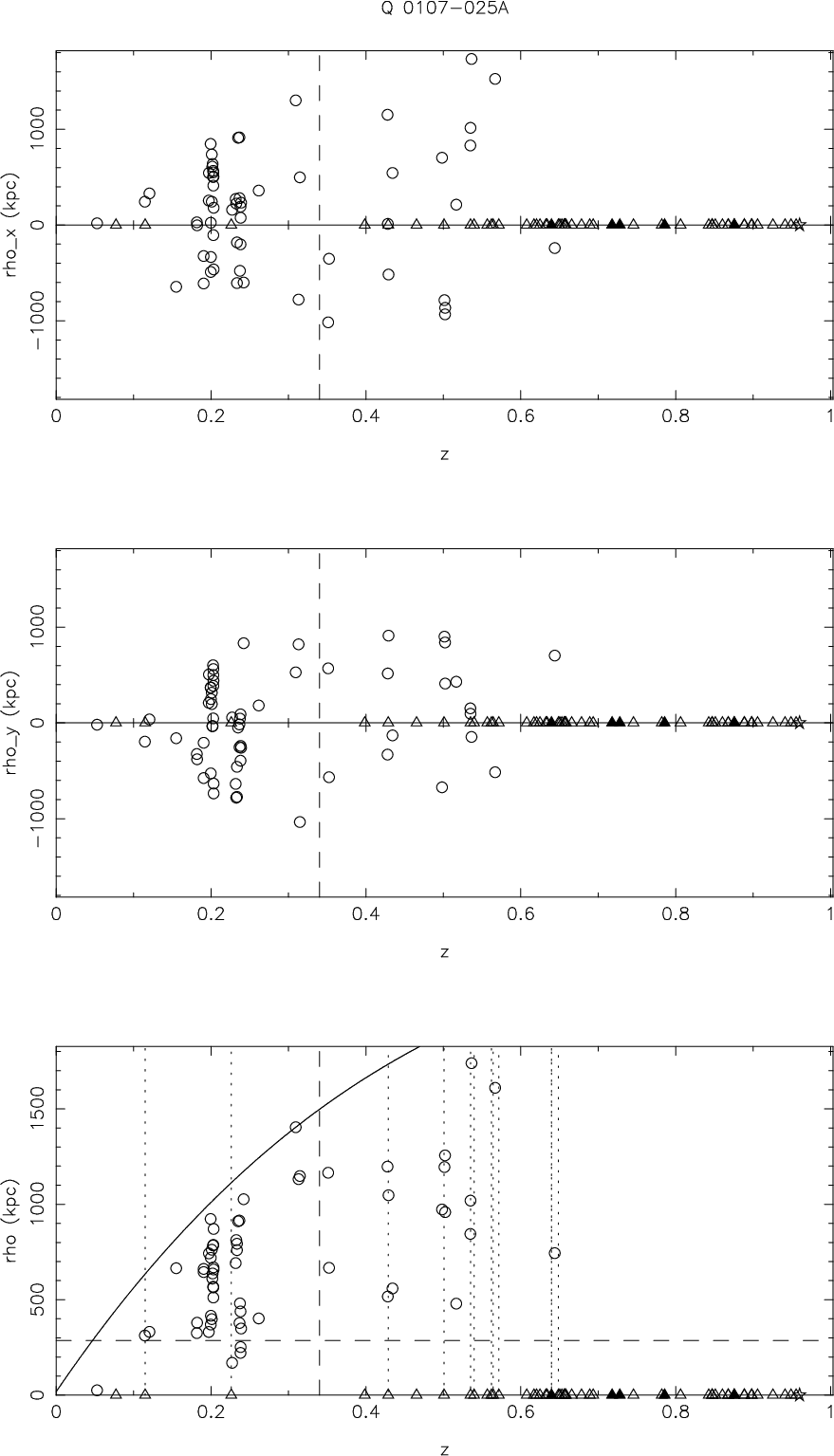}
 \caption{Pie diagram showing absorbers and galaxies in the LOS of LBQS~0107$-$025A
(includes additional objects from Rauche, see text). Description of
figure as per figure~\ref{fig-n0024pie}.}
  \label{fig-q0107apie}
\end{figure}

\begin{figure}
\includegraphics[width=84mm]{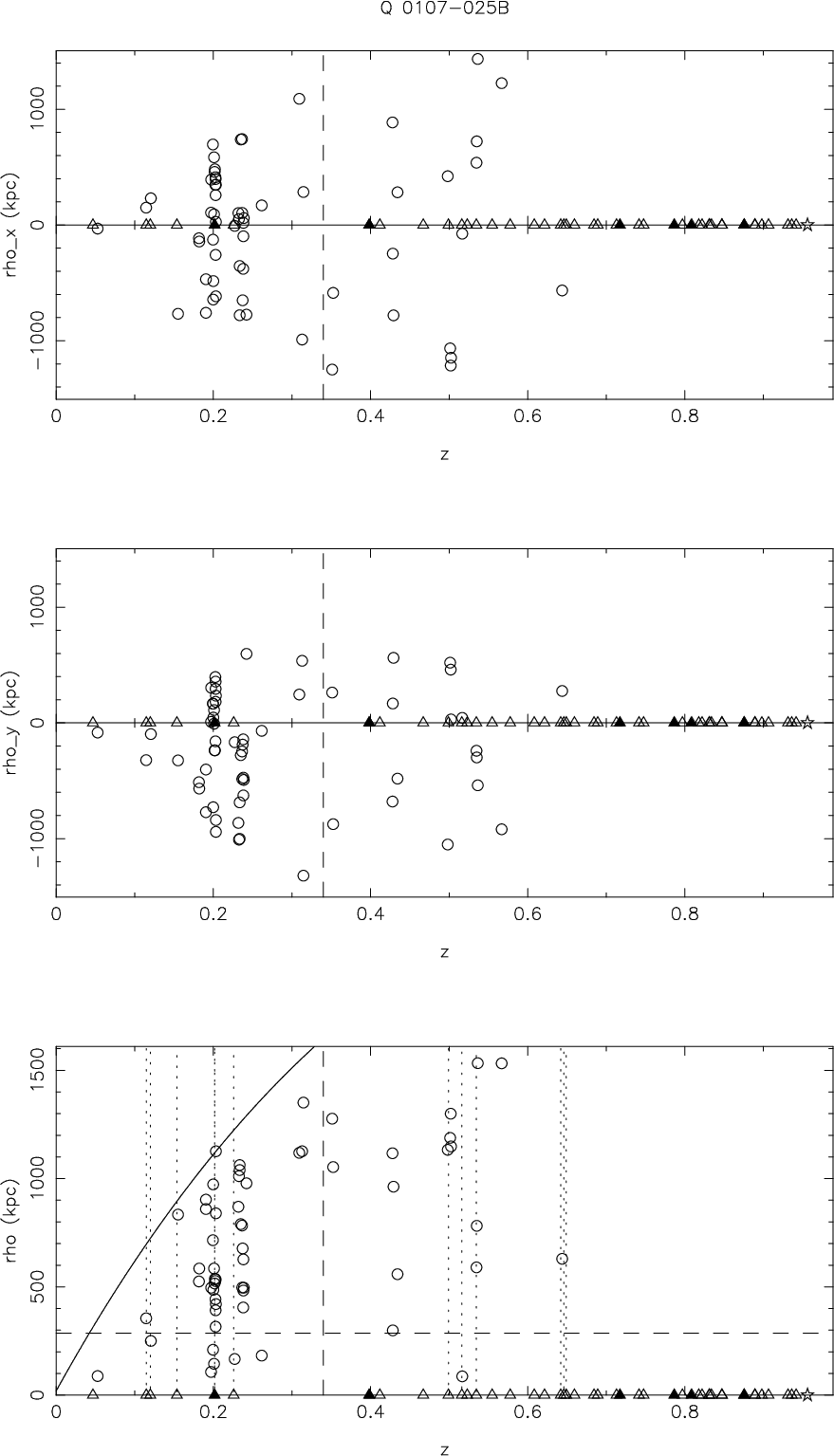}
 \caption{Pie diagram showing absorbers and galaxies in the LOS of LBQS~0107$-$025B
(includes additional objects from Rauche, see test). Description of
figure as per figure~\ref{fig-n0024pie}.}
  \label{fig-q0107bpie}
\end{figure}

\clearpage

\section{Analysis} \label{sec-anal}

\subsection{Sample Properties} \label{subsec-props}

The final galaxy sample as described in \S~\ref{sec-data} above
contains 685 objects with redshifts. Of these, 49 have `redshifts'
less than 500 \kms, and hence are almost certainly stars. Assuming
this to be true, we can check that our redshift error estimates
are reasonable. The mean velocity of this sample is 5 \kms, with
RMS 190 \kms. The mean of the error estimates on the velocities is
230 \kms, suggesting we might be slightly overestimating our
velocity errors, but, to be conservative, we make no adjustment
for this.

The absorber sample as described in \S~\ref{sec-data} above contains
815 lines, of which 381 are identified as \lya\ and 54 are CIV. Of
the CIV sample, there are 25 with unique redshifts less than one
(i.e. counting doublets as one system). For our present analysis we
will restrict ourselves to this subsample of 406 absorption lines.

\begin{figure}
\includegraphics[height=84mm, angle=90]{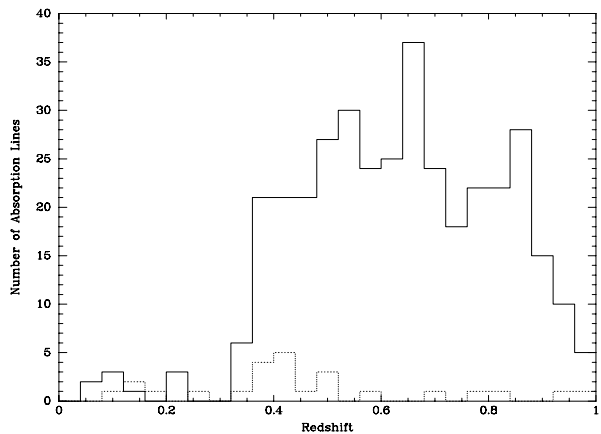}
 \caption{Absorption line sample redshift distribution.
 The solid line shows \lya\ lines, while the dotted line shows CIV.}
  \label{fig-ablineszhist}
\end{figure}

The final redshift distribution of this absorption line sample is
shown in Figure~\ref{fig-ablineszhist}. The dramatic drop in line
numbers shortward of z$\sim$0.35 is where \lya\ moves below the blue
edge of the HST FOS G190H grating. In our subsample, absorption line
systems at redshifts less than 0.35 are generally either metal line
systems, or \lya\ detected in a GHRS exposure towards LBQS0107-025A
or B.

\begin{figure}
\includegraphics[height=84mm, angle=90]{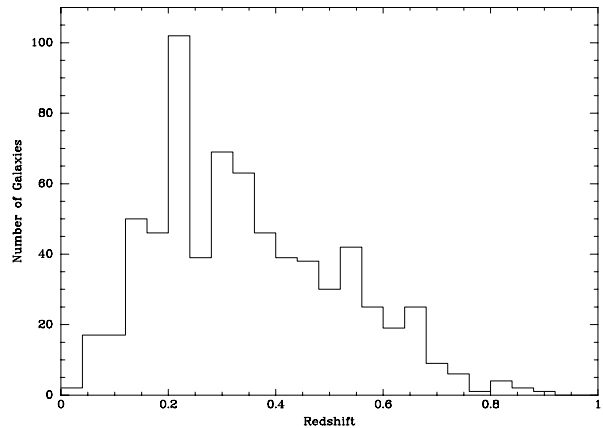}
 \caption{Galaxy Sample Redshift distribution}
  \label{fig-galzhist}
\end{figure}

The final redshift distribution of the galaxies with measured
redshifts is shown in Figure~\ref{fig-galzhist}.

\begin{figure}
\includegraphics[height=84mm, angle=90]{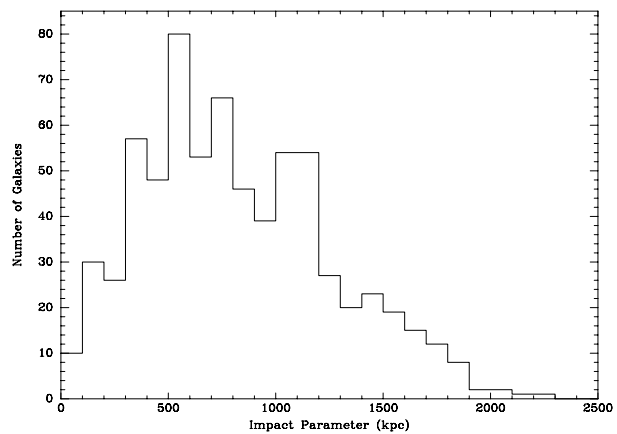}
 \caption{Galaxy Sample distribution in projected distance to QSO LOS}
  \label{fig-galphist}
\end{figure}

In Figure~\ref{fig-galphist} we show the distribution of the
galaxy sample with projected distance to the QSO LOS in our
adopted cosmology. There are relatively few galaxies within 200
kpc of the LOS (as would be expected for a uniformly sampled
distribution). The survey geometry, combined with cosmology,
results in a fairly flat distribution out to radial distances of
$\sim$1 Mpc, with a drop off from there out to 2 Mpc which
effectively defines the corners of our surveyed volume.

\begin{figure}
\includegraphics[width=84mm]{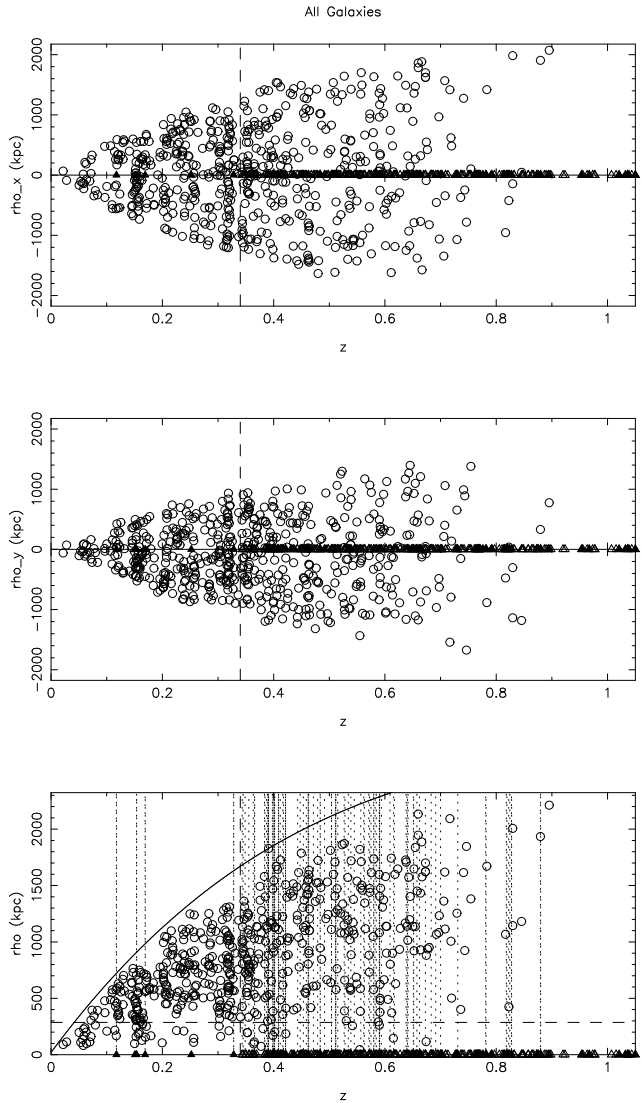}
 \caption{Pie plot showing all Lines of Sight overplotted. Description of figure
 as per figure~\ref{fig-n0024pie}. The solid curve in the lower panel shows the
 outer boundary of the survey volume. Also clearly visible is the sharp drop in
 absorber-galaxy pairs which could be found produced by the blue edge of the
 FOS G190H (corresponding to a redshift of 0.34 for \lya\ )}
  \label{fig-allpie}
\end{figure}

In Figure~\ref{fig-allpie} we show a pie diagram resulting from
combining all of the LOS. This figure should allow the reader to
understand our survey geometry and the regime over which our
statistical results apply. In particular one can clearly see that
our survey covers impact parameters out to 2 Mpc for redshifts above
0.45. One can also see that (because the KP observations did not
include wavelengths shortward of 1630{\AA}) very few absorber-galaxy
pairs are found with redshift less than 0.34. All symbols and lines
are as defined for Figure~\ref{fig-n0024pie}.

\begin{figure}
\includegraphics[height=84mm, angle=90]{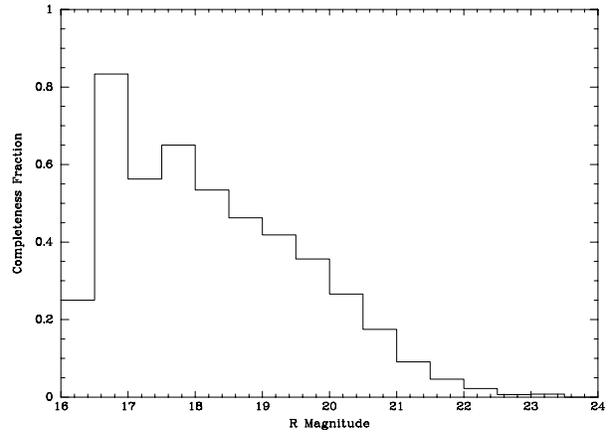}
 \caption{Galaxy sample completeness}
  \label{fig-currentcompleteness}
\end{figure}

By comparing the galaxy sample with measured redshifts with the
photometric catalogs, we can compute the completeness as a function
of magnitude for each field individually. Because of the fairly
small number statistics of individual fields, these histograms look
fairly noisy, although informative. For this paper, we show the
merged result from combining all of the fields. As can be seen in
Figure~\ref{fig-currentcompleteness}, the average completeness peaks
at around 60\% at the bright end, falling smoothly to 20\% at
R$\sim$20 and 10\% at R$\sim$21. This completeness can be taken as
roughly referring to the volume enclosed by the solid line in the
bottom panel of Figure~\ref{fig-allpie}.

\subsection{Identifying `Absorber-Galaxy pairs'} \label{subsec-pairid}

There are two reasons for trying to identify absorber-galaxy
`pairs'. One is to investigate the physical relationship (if any)
between the gas and the galaxies, while the other is to allow us
to try to tie the velocities of the galaxies to the frame of the
absorbers.

A straightforward way of tying galaxies to absorbers is to set
limits in impact parameter and velocity difference, and to call all
galaxies within that distance of an absorber `associated'. This begs
the question of what to do when there is more than one galaxy within
this distance of an absorber, with common approaches being to just
take the closest, as defined by some metric combining the velocity
difference and projected distance, or to accept all such pairs.

A more complex approach has been followed by the series of papers by
Lanzetta and Chen \citep{1995ApJ...442..538L, 1996ApJ...456L..17L,
1998ApJ...498...77C, 1999ApJ...523...72O, 2001ApJ...559..654C}, who
identify `physical' absorber-galaxy pairs by separating them from
`random' and `correlated' pairs in the following manner. `Random'
pairs are removed by calculating the value of the absorber-galaxy
correlation function assuming a parametric form (kindly provided to
us by Hsiao-Wen Chen), given the impact parameter and velocity
difference \citep{1998semi.conf..213L}. Pairs are passed to the next
stage if their calculated correlation amplitude is greater than 1.
To remove `correlated' pairs they also then require that the impact
parameter is less than 200 h$^{-1}$ kpc (286 kpc in our adopted
cosmology). Given the form of the correlation function, and the
above additional cut, one can calculate the region of impact
parameter/velocity difference space within which an absorber-galaxy
pair would be declared `physical'. We note that this is in practice
extremely similar to a simple cut at impact parameter of 286 kpc and
velocity difference of 530 \kms\ .

Another route to the above pairing was discussed by
\citet{1993ApJ...419..524M}, where the three dimensional distance
between absorbers and galaxies was corrected for the statistical
fact that objects close in projected distance might have
relatively large peculiar motions relative to each other because
they are orbiting in the same potential well. As discussed in that
paper, it is possible to use the two point correlation function to
estimate the probability distribution of the real three
dimensional separation of two objects, and then to adopt the
expectation value of this probability distribution as what was
referred to as the `perturbed Hubble flow' distance.

For simplicity we will only adopt straight cuts in impact parameter
and velocity difference (shown for example in
Figure~\ref{fig-g_a_pair_vp}), along with the Lanzetta et al.
correlation function approach, in this paper.

\subsection{Properties of $<$ 286 kpc. `Absorber-Galaxy pairs'} \label{subsec-pairprop}

In order to compare our results with those of
\citet{2001ApJ...559..654C}, we have produced pair samples using
their approach, and also with cuts in impact parameter 286 kpc and
velocity difference $\pm$530 \kms\ (the region enclosed by dashed
lines in Figure~\ref{fig-g_a_pair_vp}). For our sample, these two
approaches produce identical lists of 13 absorber-galaxy pairs.
Given our fairly large sample of \lya\ absorption lines and galaxies
with redshifts, this small number of `physical' pairs might be seen
as surprising, but we re-emphasize that our strategy in designing
the galaxy redshift survey was not to identify the galaxies
`causing' the absorption, but to obtain a statistically
well-understood sample of galaxies over a region large in impact
parameter. For this reason we did not try to concentrate our efforts
at small impact parameters. Indeed, this would have been a rather
inefficient use of telescope time and MOS spectrograph real-estate.

\begin{figure}
\includegraphics[height=65mm, angle=0]{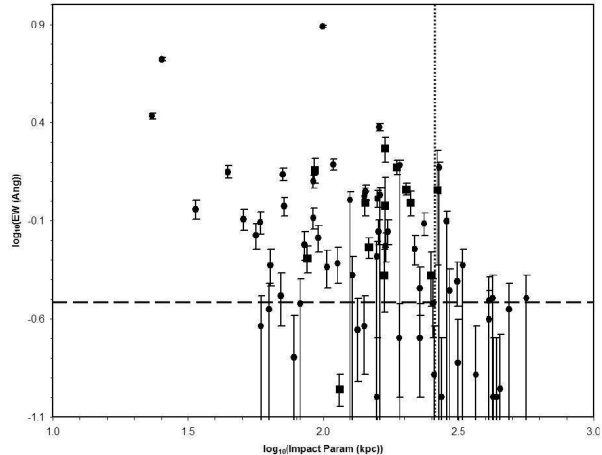}
 \caption{Relationship between Equivalent width and impact parameter for pairs of absorbers
   seen in \lya\ and galaxies selected as described in the text. Small circles show the
   data of \citet{2001ApJ...559..654C}, while larger squares show the new pairs from this paper.
   Also marked are the fiducial EW of 0.3{\AA} and impact parameter of 257 kpc (see text)}
  \label{fig-chen_fig3}
\end{figure}

\citet{2001ApJ...559..654C} use their sample of 34 galaxy-absorber
pairs to conclude that L* galaxies have tenuous gaseous halos of
column density $\ge$10$^{14}$ cm$^{-2}$, with radius 180 h$^{-1}$
kpc and covering factor near unity, and that the halo radius scales
with galaxy B-band luminosity as L$_{\rm B}^{0.39}$. Their Figures~3
and 4 make the case for this conclusion. In
Figure~\ref{fig-chen_fig3} we show a reproduction of their Figure~3
(converted to our adopted cosmology) along with the additional pairs
from our sample. Also marked are the fiducial EW of 0.3{\AA}
(corresponding roughly to a column density of 10$^{14}$ cm$^{-2}$)
and the L* impact parameter of 257 kpc (180 h$^{-1}$ kpc with
h=0.7).

In response to an interesting question from the referee of this
paper, we have investigated the sub-sample of pairs found with large
rest EW (taken to mean larger than 1{\AA}) and large impact
parameter (taken to mean larger than ~100 kpc). There are 5 such
pairs in our sample. With such a small sample, it is impossible to
draw any statistically convincing conclusions, but it is notable
that all 5 galaxies so identified have at least one other galaxy
close to it in redshift. The clearest example is the absorber galaxy
pair in the PG~2302+029 LOS at redshift 0.59 where there is a clear
group or wall of galaxies at this redshift (see
Figure~\ref{fig-p2302pie}). One possible inference, which could only
be confirmed with a much larger sample, would be that large EW,
large impact parameter `pairs' are identified in somewhat overdense
regions.

The additional pairs from our sample do significantly increase the
number of large column density pairs with separations larger than
100 kpc, but the incomplete nature of our sample (and indeed of most
samples of this nature) does leave open the possibility that a
galaxy of comparable brightness might be found closer to the LOS.

We also looked at the rest frame B-band luminosities of the galaxies
that are paired in this manner, finding that these were not
significantly different from the luminosities of all the galaxies
found in pairs with separation up to $\pm$5000 \kms\ and impact
parameter up to 2 Mpc. I.e. there is no evidence from our sample
that more luminous galaxies are more likely to be found in
absorber-galaxy pairs. However, the large uncertainties in the
derived absolute B-band magnitudes we tabulate, that results largely
from an uncertain k-correction, mean that any such trend, even if
present, could be undetectable using the data in this paper.

As a result of these inconclusive results, we have chosen to abandon
the approach of trying to identify a single `physical'
galaxy-absorber pair, and in the next section will concentrate on
the statistical properties of all the galaxy-absorber pairs within a
fairly generous window.

\subsection{Properties of All Absorber$-$Galaxy Pairs with velocity difference
less than 1000 \kms} \label{subsec-monte}

\begin{figure}
\includegraphics[height=84mm, angle=90]{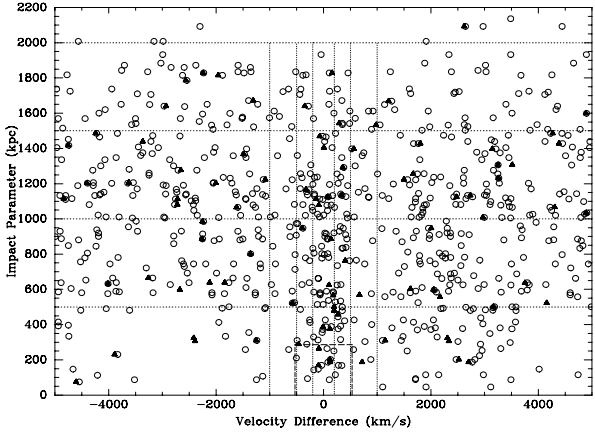}
 \caption{Distribution of velocity differences and impact parameter between absorbers
   seen in \lya\ or CIV and
   galaxies selected as described in the text. Absorbers
   seen in \lya\ only are marked by open circles, absorbers seen in CIV are marked as filled circles.}
  \label{fig-g_a_pair_vp}
\end{figure}

We can now analyse the statistical properties of the relative
distributions of the gas and galaxies along the 16 lines of sight.
The basic sample we will use will be taken from all possible pairs
of absorbers and galaxies with velocity differences less than +/-
5000 \kms\ .  For the CIV systems we have only taken the stronger
line of the doublet. We note that therefore individual absorbers or
galaxies can appear multiple time as different pairs. The entire
sample is shown in Figure~\ref{fig-g_a_pair_vp}. This complicated
plot shows both the regions we have determined to contain an excess
of absorber-galaxy `pairs', and also a region beyond +/- 1000 \kms\
which we will use to determine the `background' value for random
pairings. Our sample is large enough to break into bins of impact
parameter (shown divided by horizontal lines in the figure). We also
measure absorber-galaxy pair over-densities for three different
velocity ranges +/- 200, 500 and 1000 \kms\ . We show the
illustrative box at impact parameter of 286 kpc and velocity
difference of 530 \kms\ which approximates the region selected by
\citet{1995ApJ...442..538L, 1996ApJ...456L..17L,
1998ApJ...498...77C, 1999ApJ...523...72O} and
\citet{2001ApJ...559..654C} as containing physical pairs.

\begin{figure}
\includegraphics[height=84mm, angle=90]{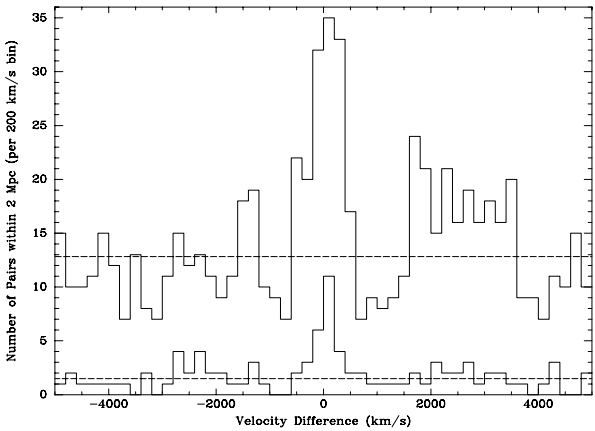}
 \caption{Distribution of velocity differences between absorbers seen in \lya\ or CIV
     with galaxies selected as described in the text. The upper line shows the pairs with absorbers
   seen in \lya\ only while the lower line shows pairs with absorbers seen in CIV.}
  \label{fig-pairvdiffboth}
\end{figure}

In Figure~\ref{fig-pairvdiffboth} we show the above data projected
onto the x-axis and displayed as a histogram. We also show the mean
background values for the \lya\ and CIV samples. Two thing can be
readily seen from this figure. First that there is a strong
over-density of pairswith a small velocity difference in both \lya\
and CIV, and secondly that the velocity differences seen in CIV are
somewhat smaller than those seen in the \lya\ . The RMS of the
velocity differences between all absorber-galaxy pairs seen in \lya\
with velocity difference between $\pm$1000 \kms\ is 440 \kms\ (191
pairs), while the RMS for absorber-galaxy pairs seen in CIV with
velocity difference between $\pm$1000 \kms\ is 350 \kms\ (31 pairs).

\begin{figure}
\includegraphics[height=84mm, angle=90]{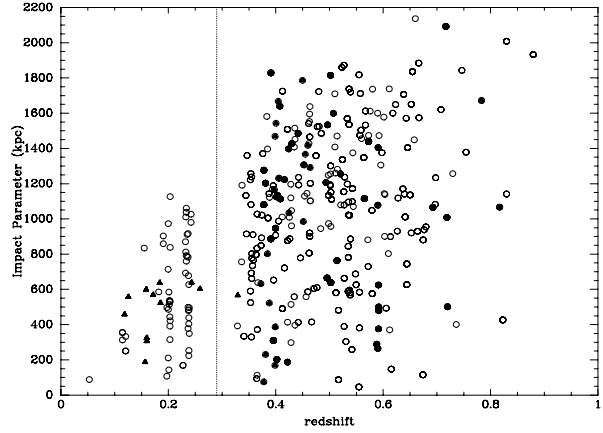}
 \caption{Distribution of redshift and impact parameter for absorber-galaxy pairs. Pairs with absorbers
   seen in \lya\ only are marked by open circles, pairs with absorbers seen in CIV are marked as filled triangles. }
  \label{fig-g_a_pair_zp}
\end{figure}

We also show in Figure~\ref{fig-g_a_pair_zp} how the pairs in the
two other plots are distributed in redshift. As a result of the
wavelength coverage of the FOS spectra providing our absorption line
sample, there is a clear break in the sample at z=0.29. For the
following analysis we will consider both the whole sample, and also
just the pairs with z$>$0.29 to check whether the results are being
affected by the low redshift sub-sample (which mostly comes from the
LBQS 0107-025A and B LOS).

\begin{figure}
\includegraphics[height=60mm, angle=0]{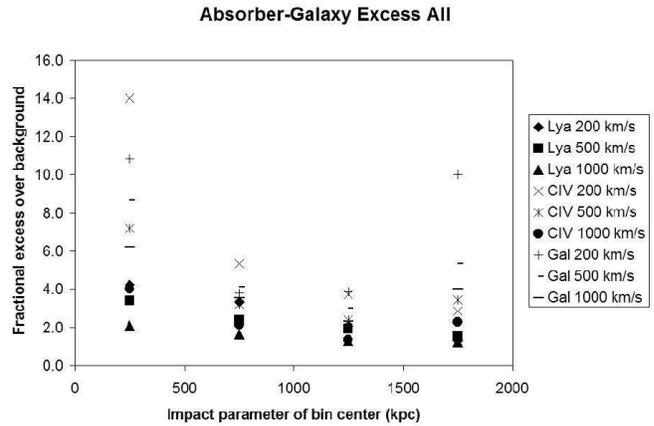}
 \caption{Measured Excess of Absorber-Galaxy Pairs relative to the Background }
  \label{fig-a_g_fr_all}
\end{figure}

\begin{figure}
\includegraphics[height=60mm, angle=0]{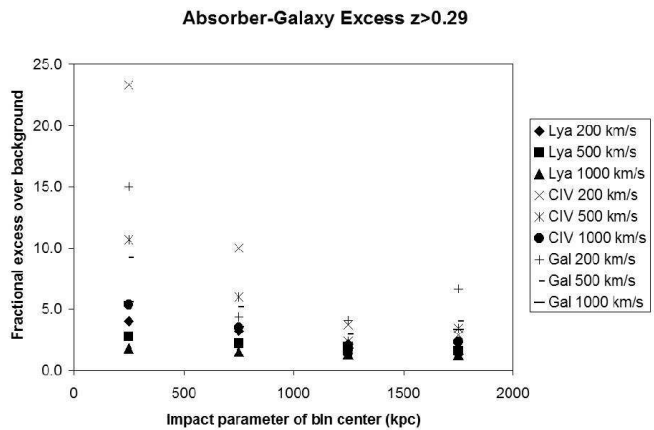}
 \caption{Measured Excess of Absorber-Galaxy Pairs with z$>$0.29 relative to the Background }
  \label{fig-a_g_fr_hz}
\end{figure}

\begin{figure}
\includegraphics[height=60mm, angle=0]{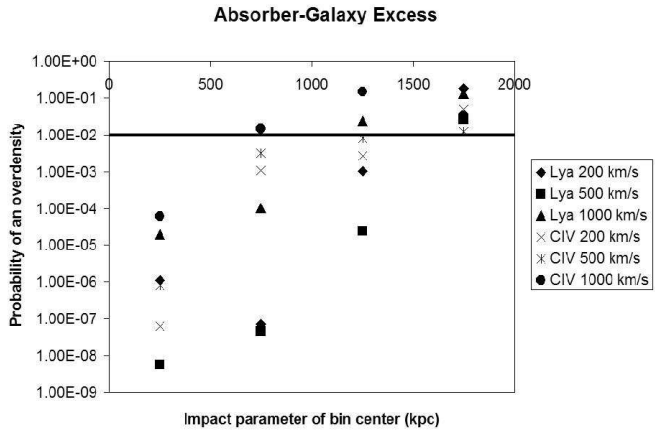}
 \caption{Significance level of the Measured Excess of Absorber-Galaxy
 Pairs relative to the Background. The number plotted on the y-axis is the probability
 that the number of pairs found within velocity range given in the sidebar
 would be found given the mean background and no real excess.}
  \label{fig-a_g_prob_all}
\end{figure}

\begin{figure}
\includegraphics[height=60mm, angle=0]{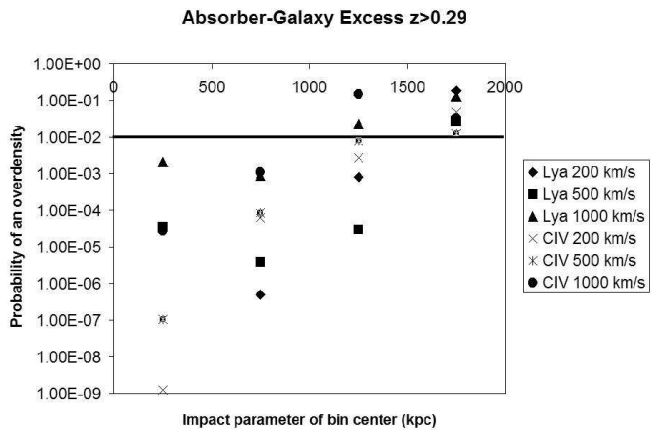}
 \caption{Significance level of the Measured Excess of Absorber-Galaxy Pairs with z$>$0.29 relative
 to the Background. The number plotted on the y-axis is the probability
 that the number of pairs found within velocity range given in the sidebar
 would be found given the mean background and no real excess.}
  \label{fig-a_g_prob_hz}
\end{figure}

Given this sample, we can now pose the question: How large is the
absorber-galaxy pair over-density compared to the background, and
how significant is the detection of this over-density?  In order to
measure the latter, we consider the numbers of pairs seen in the
various bins in impact parameter and velocity difference in
Figure~\ref{fig-g_a_pair_vp} to be drawn from a Poisson distribution
with an expected mean given by the `background' density measured at
the same range of impact parameter but with velocity differences
between -5000 and -1000 \kms\ and between +1000 and +5000 \kms\ .
The results of the analysis are shown in
Figures~\ref{fig-a_g_fr_all}, \ref{fig-a_g_fr_hz},
\ref{fig-a_g_prob_all}, and \ref{fig-a_g_prob_hz}.

In Figures~\ref{fig-a_g_fr_all} and \ref{fig-a_g_fr_hz} we show that (as
expected by a visual inspection of Figure~\ref{fig-g_a_pair_vp})
there is a substantial over-density of absorber-galaxy pairs at
impact parameters less than 500 kpc. The fractional over-density is
larger for the systems seen in CIV than for the \lya\ systems. At
larger impact parameters, the fractional over-density is less. It
is reasonable that the strength of the over-density generally drops
as one increases the velocity difference range within which one
counts pairs.

In Figures~\ref{fig-a_g_prob_all} and \ref{fig-a_g_prob_hz} we show
that the over-densities notable in Figures~\ref{fig-a_g_fr_all} and
\ref{fig-a_g_fr_hz} have a $<$1\% probability of arising by chance
out to impact parameters of at least 1.5 Mpc. This is consistent,
for example, with the models of \citet{1999ApJ...511..521D}.
Figure~18 in that paper for example shows correlations in
absorber-galaxy properties out to these sorts of scales.

Note that the probabilities reported assume that the expected (or
background) value is known perfectly. In practice there is a
(Poisson) error on our measurement of this background, which could
be folded in to the analysis, but has not been in order to avoid
overly complicated plots.

As noted in \S~\ref{subsec-kpabs}, the identifications of the lines
in the PG 1718+481 LOS are not complete. We have rerun the above
statistical tests including all the unidentified lines as \lya\  and
find that this makes no significant difference to the results
reported in this section.

We also remind the reader that the FOS spectral resolution of 230
\kms\ means that no pairs of absorption features can be identified
with velocity difference smaller than this. The low density of \lya\
absorbers at these redshifts make the likelihood of two such lines
coinciding by chance low, although clustering can obviously change
this conclusion. This issue is discussed in more detail in
\citet{1998ApJS..118....1J}. We also note that the galaxy redshift
uncertainties (also of ~230 \kms\ ) will artificially move galaxies
across the boundaries of any velocity difference cut. The general
trend will be (a) to weaken any measured correlation, and (b) for
this weakening to be stronger for the smaller velocity difference
bins.

\begin{figure}
\includegraphics[height=84mm, angle=90]{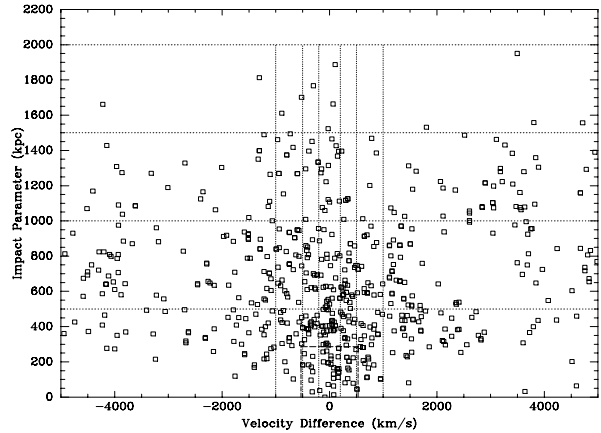}
 \caption{Distribution of velocity difference and impact parameter between galaxies and
   galaxies selected as described in the text. }
  \label{fig-g_g_pair_vp}
\end{figure}

\begin{figure}
\includegraphics[height=84mm, angle=90]{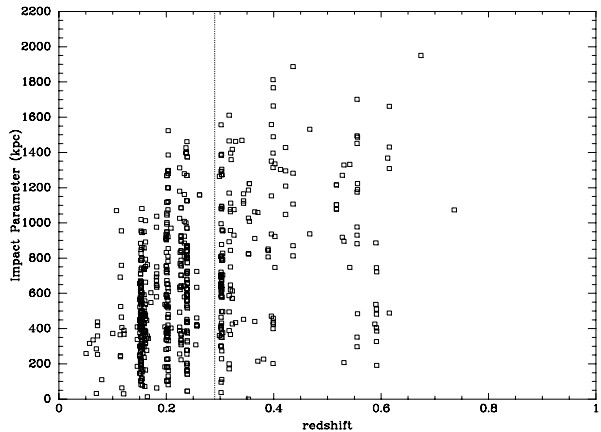}
 \caption{Distribution of redshift and impact parameter between galaxies and
   galaxies selected as described in the text. }
  \label{fig-g_g_pair_zp}
\end{figure}

An obvious question is whether the observed results are simply due
to the absorbing gas being in the halo of a galaxy, and the detected
pair excess being due to the galaxy-galaxy correlation function.
This is tested using the galaxy sample and forming a similar sample
of galaxy-galaxy pairs to the absorber-galaxy pairs. In order to
approximately match the absorber galaxy sample, we choose galaxies
from each field within 500 kpc of the QSO LOS, and collect up all
resulting pairs with that sample. This subsample should then be able
to find pairs in the same volume of space as the absorber sample.
Figures~\ref{fig-g_g_pair_vp} and \ref{fig-g_g_pair_zp} are the
equivalent of Figures~\ref{fig-g_a_pair_vp} and
\ref{fig-g_a_pair_zp} but for the galaxy-galaxy pairs. The
over-densities found for galaxy-galaxy pairs were plotted in
Figures~\ref{fig-a_g_fr_all} and \ref{fig-a_g_fr_hz}.

The above analysis shows that the galaxy-galaxy pair over-density is
significantly higher than the absorber-galaxy pair over-density seen
for the \lya\ systems, but comparable to that seen for the absorber
systems seen in CIV. This result could have several interpretations,
but one might be that the absorbing gas seen in CIV (and which has
obviously been polluted by outflows from galaxies) is still in close
proximity with galaxies, while systems seen in \lya\ are more widely
distributed, albeit still correlated with the galaxy
distribution.\footnote{When this paper was near to submission, an
interesting short letter appeared on astro-ph by
\citet{2005ApJ...629L..25C} describing a single LOS with high
quality STIS UV spectroscopy. They perform a correlation analysis
with a sample of 61 galaxies within 1 h$^{-1}$ Mpc at redshifts less
than 0.5. Some of their conclusions will need confirmation with a
greater number of LOS and a larger galaxy sample, but their
conclusion that the absorber-galaxy correlation is entirely due to
the emission line (i.e. star forming) galaxies is potentially very
exciting. If confirmed, then this would support a seductively
simplistic model with star forming galaxies and high column density
absorbing gas co-existing in the filamentary structure now so
familiar from the simulations, while the more highly clustered
absorption line galaxies lie in the knots and intersections where
the 10$^{4}$ K gas needed for detection in \lya\ is absent or
destroyed (possibly by conversion into the much sought after WHIM).
Our data set can be used to look for confirmation of this picture.}
Another possible interpretation is that we are just seeing the
column density correlation with impact parameter noted by other
authors and modelled for example by \citet{1999ApJ...511..521D}.
Separating hydrogen column density from metallicity as the
underlying driver of these correlations needs further work and will
benefit from higher resolution UV spectroscopy (i.e. data from the
HST archive taken with the Space Telescope Imaging Spectrograph
(STIS)).

\subsection{The link to simulations and theory} \label{subsec-sims}

One of the many benefits provided by the powerful cosmological
simulations begun in the 1990s and continuing into the present (see
introduction for list of references),  is the ability to simulate
the evolution of the physical properties of \lya\ absorbers as a
function of redshift and cosmic time. This helps tremendously when
trying to make use of diverse observational data sets in order to
understand what physical processes influence the physical properties
and evolution of the IGM and galaxies. The theoretical models aid us
in identifying the high redshift progenitors of structures observed
at low redshift. For example, it is from such models (e.g.
\citet{1999ApJ...511..521D, 2001ApJ...559..507S}) that we understand
that a given relative over-density in the (dark) matter distribution
will be associated with a different column density of neutral
hydrogen as a function of redshift.  \lya\ absorbers with \nhi\
approximately 10$^{14}$ cm$^{-2}$ at $z=0$ are in regions of
over-density that at $z=3$ would have neutral gas columns closer to
10$^{16}$ cm$^{-2}$ \citep{1999ApJ...511..521D}. We note that this
does not imply that regions of neutral column density 10$^{16}$
cm$^{-2}$ at z=3 {\bf evolve} into regions with neutral column
density 10$^{14}$ cm$^{-2}$ at z=0, but merely that these two
different column densities of neutral material flag regions at those
two epochs with the same dark matter overdensity.

A natural next step for this work will be to compare the statistics
we have derived from our observations with statistics measured as
close as possibly identically on simulations which include feedback,
with a goal of verifying or ruling out the feedback assumptions they
have made. This is a challenging project, as the simulation must
faithfully include both low density inter-galactic gas, and also
accurately predict the locations and physical properties (stellar
luminosities and star formation histories) of the highly-collapsed
galaxies. The redshift range from z=1 to the present day is well
known to be hard to model at high resolution, because of the large
dynamic range needed. That said, several of the references to
modelling work given in the introduction suggest that this problem
is tractable.

\section{Conclusions} \label{sec-conc}

A goal of our work is an improved understanding of how gas is
transformed into galaxies, and how those galaxies in turn influence
the gas around them (e.g. by winds, jets or radiation).  Studies of
absorbers and galaxies at high redshift have already yielded
evidence for such mechanisms playing a role in the formation and
evolution of galaxies (e.g. \cite{2003ApJ...584...45A};
\citet{2002ApJ...580..634C}; \citet{2003A&A...407..473F};
\citet{2003MNRAS.343L..41B} and \citet{2003ApJ...594...75K}).

We have focussed on the second half of the history of the Universe,
i.e. at redshifts less than one. This is a period in which the star
formation density of the Universe is thought to be dropping away
from its peak, the rapid evolution of the properties of the \lya\
absorbers has slowed, and when the Hubble sequence of galaxies is
well established. Despite this apparent middle-aged placidity, there
should remain clear evidence of the wild excesses of youth, and that
is what we claim to have measured. An advantage of focussing on this
(large) fraction of the history of the Universe is that the galaxy
population can be studied in detail, and with confidence that a
representative sample has been obtained.

There is evidence from the KP data set alone for clustering of some
fraction of low redshift ($z<1$)  \lya\ absorbers on velocity scales
$<$ 300 km/sec. This evidence is that approximately 10\% of the KP
\lya\ absorbers are ``resolved'' in the FOS spectra
\citep{1998ApJS..118....1J}.

At higher column densities, the Lyman limit systems contained in the
KP sample all have extensive associated metal line systems.  For all
the extensive metal line systems (multiple absorption lines from
strong resonance line producing species, e.g. Si, O, N, etc.) for
which we could check for the presence of an associated Lyman Limit
system, we detect such a system. In other words, the absorption line
systems that are unambiguously higher column density systems (at low
redshift) are unquestionably more like the gas producing the ISM of
a galaxy than like the gas one might expect in a void. These same
absorbers also often have associated OVI, or NV, or other evidence
of more highly ionized components i.e. multi-phase gas again, like a
galaxy.

Our conclusions for the more moderate column density systems that
dominate our current sample can be summarized as:

\begin{itemize}
\item A correlation between absorbers and galaxies has been detected
out to impact parameters of at least 1.5 Mpc.

\item The strength of the absorber-galaxy correlation is weaker than
the correlation between galaxies and galaxies.

\item The velocity differences seen between galaxies and absorbers
with detected CIV are typically smaller than the velocity
differences seen between galaxies and absorbers seen in \lya\ only.

\item The above is consistent with absorbers being a mixed population
containing some 'pristine'\footnote{Given the results of e.g.
\citet{2001ApJ...561L.153S} on the high-z Universe that show that
all gas has CIV by z of 5.5, it is debatable whether anything we see
in absorption is truly `pristine'} material probably infalling for
the first time along the filaments predicted by current models, and
some 'contaminated' material produced by outflows which in general
lies closer to galaxies in velocity.

\item The above qualitative picture needs to be fleshed out by comparing
the numerical strengths of the observed correlations and
relationships with SPH or AMR modelling.
\end{itemize}

While we have so far only made a fairly superficial analysis of the
(no-doubt complex) relationship between gas and galaxies at
redshifts less than one, we feel the above results are very
encouraging, and that they suggest that there is considerable room
for further observation and modelling of this crucial interaction.

\section*{Acknowledgments}

This work would not have been possible without many landmark
contributions by Professor John Bahcall, who sadly passed away as we
prepared to submit this paper. Not only did Bahcall anticipate the
production of quasar absorption lines by gas associated with
galaxies and other structures \citep{1969ApJ...156L..63B}, but he
indefatigably worked for the development, launch, repair, and
continued operation of the Hubble Space Telescope. As PI of the Key
Project, he led the team that successfully completed the first
census of absorbers at low redshift, enabling the present work.

We would like to that our anonymous referee who made a number of
suggestions that significantly improved the quality of this paper.

We thank Ray Weymann for his active participation in the hard work
of the observing runs that produced the data presented in this work
and for many relevant conversations. We also thank the CFHT
telescope operators. We thank Michael Rauch for providing the
additional redshifts in the LBQS~0107$-$025 sight-line and H. W.
Chen for providing to us the description of the correlation function
used in the referenced Chen et al. papers. We would like to
acknowledge valuable discussions regarding the proper use of FOCAS
and SExtractor with respectively F. Valdes and M. J. I. Brown. We
also thank Professor Carlos Frenk and the Durham visitor grant
program that helped support this work. BTJ acknowledges partial
support of early stages of this work from NASA through
HF-1045.01-93A and later stages by the National Optical Astronomy
Observatory, operated by AURA, Inc., on behalf of the U.S. National
Science Foundation. Finally, we thank Mrs.~Jane Darwin for providing
the accommodation where the production of this document was finally
completed.

\bibliographystyle{mn2e}
\bibliography{mn-jour,refs_mj_cfht1_16_11_05}

\bsp

\label{lastpage}

\end{document}